\documentclass{article}

\usepackage{PRIMEarxiv}

\usepackage[utf8]{inputenc} 
\usepackage[T1]{fontenc}    
\usepackage{hyperref}       
\usepackage{url}            
\usepackage{booktabs}       
\usepackage{amsfonts}       
\usepackage{nicefrac}       
\usepackage{microtype}      
\usepackage{lipsum}
\usepackage{fancyhdr}       
\usepackage[dvips]{graphicx}
\usepackage{graphicx}       
\graphicspath{{media/}}     

\usepackage{amsmath}
\usepackage{graphics}
\usepackage{amssymb}
\usepackage[pdftex]{color}
\usepackage{tabularx}
\usepackage{longtable}
\usepackage{pdflscape}
\usepackage{multirow}
\usepackage{booktabs}
\usepackage{setspace}
\usepackage{natbib}
\usepackage[ruled,vlined]{algorithm2e}
\usepackage{multirow}
\usepackage{enumitem}

\usepackage{listings}
\usepackage{xcolor}
\definecolor{mauve}{RGB}{136,0,21}
\definecolor{dkgreen}{rgb}{0,0.6,0}
\lstdefinestyle{mystyle}{
    language=R,
    commentstyle=\color{dkgreen},
    keywordstyle=\color{blue},
    numberstyle=\color{gray},
    stringstyle=\color{mauve},
    basicstyle=\ttfamily\footnotesize,
    breakatwhitespace=false,         
    breaklines=true,                 
    captionpos=b,                    
    keepspaces=true,                 
    showspaces=false,                
    showstringspaces=false,
    showtabs=false,                  
    tabsize=2
}
\allowdisplaybreaks

\newtheorem{theorem}{Theorem}[section]
\newtheorem{proposition}[theorem]{Proposition}

\newtheorem{remark}[theorem]{Remark}

\newcommand{\X}{\textbf{X}}
\newcommand{\Y}{\textbf{Y}}
\newcommand{\Z}{\textbf{Z}}
\newcommand{\x}{\textbf{x}}
\newcommand{\y}{\textbf{y}}

\newcommand{\as}{\textbf{s}}

\newcommand{\N}{\mathcal{N}}
\newcommand{\E}{\mathbb{E}}
\newcommand{\bmu}{\boldsymbol{\mu}}
\newcommand{\bbeta}{\boldsymbol{\beta}}
\newcommand{\bSigma}{\boldsymbol{\Sigma}}
\newcommand{\bLambda}{\boldsymbol{\Lambda}}

\newcommand{\bOmega}{\boldsymbol{\Omega}}
\newcommand{\btheta}{\boldsymbol{\theta}}
\newcommand{\bPsi}{\boldsymbol{\Psi}}
\newcommand{\balpha}{\boldsymbol{\alpha}}
\newcommand{\bxi}{\boldsymbol{\xi}}

\newcommand{\amax}[1]{\underset{#1}{\operatorname{argmax}}}

\newcommand{\dr}{\mathrm{d}}

\title{Moments and Random Number Generation for the Truncated Elliptical Family of Distributions
}

\author{
  Katherine A. L. Valeriano \\
  Departamento de Estatística \\
  Universidade Estadual de Campinas \\
  Campinas, Brazil 13083-970\\
  \texttt{katandreina@gmail.com} \\
  \And
  Christian E. Galarza \\
  Departamento de Matemáticas \\
  Escuela Superior Politécnica del Litoral \\
  Guayaquil, Ecuador 090112\\
  \texttt{chedgala@espol.edu.ec} \\
  \And
  Larissa A. Matos \\
  Departamento de Estatística \\
  Universidade Estadual de Campinas \\
  Campinas, Brazil 13083-970\\
  \texttt{larissam@unicamp.br} \\
}

\begin{document}
\maketitle

\begin{abstract}
This paper proposes an algorithm to generate random numbers from any member of the truncated multivariate elliptical family of distributions with a strictly decreasing density generating function. Based on \cite{neal2003slice} and \cite{ho2012some}, we construct an efficient sampling method by means of a slice sampling algorithm with Gibbs sampler steps. We also provide a faster approach to approximate the first and the second moment for the truncated multivariate elliptical distributions where Monte Carlo integration is used for the truncated partition, and explicit expressions for the non-truncated part \citep{galarza2020moments}. Examples and an application to environmental spatial data illustrate its usefulness. Methods are available for free in the new  \textsf{R} library \textbf{relliptical}.
\end{abstract}

\keywords{Elliptical distributions \and Slice sampling algorithm \and Truncated distributions \and Truncated moments}

\section{Introduction}
The use of truncated distributions arises in a wide variety of statistical models as survival analysis, censored data models, Bayesian models with truncated parameters space, and abound in such fields as agronomy, biology, environmental monitoring, medicine, and economics. Algorithms like Expectation-Maximization (EM) \citep{dempster1977maximum} are employed frequently in multivariate censored data analysis under a likelihood-based perspective for its facility to deal with missing and partially observed data. This algorithm requires the computation of conditional truncated moments, commonly the first two moments. For example, \cite{matos2013likelihood} and \cite{matos2016censored} estimated the parameters of a censored mixed-effects model for irregularly repeated measures via the EM algorithm, which needed to compute the first two moments of a truncated multivariate $t$ (TMVT) and a truncated multivariate normal (TMVN) distributions, respectively. 

In this context, there are a few libraries in  \textsf{R} \citep{Rproj} which provide truncated multivariate moments. For instance, the package \textbf{tmvtnorm} \citep{wilhelm2015package} computes the mean and the variance of the TMVN distribution by deriving its moment generating function, while the \textbf{MomTrunc} library \citep{galarza2021package} uses a recursive approach method proposed by \cite{kan2017moments} to compute arbitrary higher-order moments. On the other hand, for the TMVT distribution, the packages \textbf{TTmoment} \citep{ho2015r} and \textbf{MomTrunc} compute its two first moments. Moreover, the first library only handles integer degrees of freedom greater than 4, while the latter can compute even high-order moments for any degrees of freedom \citep{galarza2021moments}.

Variations of the EM algorithm such as Stochastic Approximation EM (SAEM) \citep{delyon1999convergence} and Monte Carlo EM (MCEM) \citep{wei1990monte} replace the conditional expectations by an approximation that requires to draw independent random observations from a truncated distribution. For instance, \cite{lachos2017influence} estimated the parameters of a linear spatial model for censored data using the SAEM algorithm, which needed to generate random samples from the TMVN distribution to perform the stochastic approximation step.
More recently, also using the SAEM algorithm, \cite{lachos2019flexible} proposed a robust multivariate linear mixed model for multiple censored responses based on the scale mixtures of normal (SMN) distributions. Moreover, generating random numbers from truncated distributions is also required in Bayesian models, \cite{gelfand1992bayesian} showed how to perform Bayesian analysis for constrained parameters or truncated data problems by using Gibbs sampling. 

There are several methods to generate random samples from a truncated distribution in the literature, and the common one is the rejection sampling technique. This method draws samples from the non-truncated distribution and retains only the samples inside the support region. However, the procedure may be inefficient, especially when the truncation interval is too small or it is located at a less probable area of the probability density function (pdf). \cite{neal2003slice} proposed  proposed the Slice sampling method, a procedure that turns sampling from a truncated density into sampling repeatedly from uniform distributions instead. This algorithm is easy to code, fast and does not reject samples, making it more efficient than the conventional rejection method.

To the best of our knowledge, there are no proposals in the literature to generate samples from other
multivariate truncated distributions in the elliptical class other than the TMVN and TMVT distributions
(available in the \textbf{tmvtnorm} and \textbf{TTmoment} packages). Hence, motivated by the slice sampling algorithm, we propose a general method to obtain samples from any truncated multivariate elliptical distribution with strictly decreasing density generating function (dgf). Using conditional expectation properties, we also propose an efficient algorithm to approximate the moments of the most common distribution of this class: the truncated multivariate
normal, Student-$t$, slash, contaminated normal, and Pearson VII distributions. This method requires less running time when compared with the existing ones, since it deals with the truncated and non-truncated part of the vector separately. Our proposal can be reached through the \textsf{R} package \textbf{relliptical}. Finally, it is worth mentioning that moments of truncated elliptical distributions can be used to compute truncated moments for the selection elliptical family of distributions, a wide family which includes complex multivariate asymmetric versions of the elliptical distributions as the extended skew-normal, the unified skew-$t$ distributions, among others. Therefore, our proposal opens the doors for the calculation of truncated moments of complex elliptical asymmetric distributions, which are of particular interest for the development of robust censored models with asymmetry, heavy tails and missingness \citep[see for instance][]{galarza2021skew, de2021finite}.

The paper is organized as follows. Section \ref{preliminaries} shows some results related to the elliptical and truncated elliptical family of distributions and a brief description of the slice sampling algorithm. Section \ref{sliceSampling} is devoted to the formulation of the sampling algorithm for the truncated elliptical distributions, whereas Section \ref{moments} focuses on our proposed method to approximate the first and the second moment. For the last two sections, we present a brief introduction to its respective  \textsf{R} function. A simulation study that compares the mean and covariance matrix for the TMVT distribution estimated through different methods in  \textsf{R} is presented as well. Section \ref{application} displays an application on censored Gaussian spatial models throughout the analysis of the Missouri dioxin contamination dataset. Finally, Section \ref{conclusions} concludes with a discussion.


\section{Preliminaries}\label{preliminaries}

\subsection{Elliptical Family of Distributions}
As defined in \cite{muirhead2009aspects} and \cite{fang2018symmetric}, a random vector $\X\in\mathbb{R}^p$ is said to follow an elliptical distribution with location parameter $\bmu\in\mathbb{R}^p$, positive-definite scale matrix $\bSigma\in\mathbb{R}^{p\times p}$, and density generating function $g$, if its pdf is given by
\begin{equation}\label{elliptical}
f_\X(\x) = c_p |\bSigma|^{-1/2} g\left( (\x - \bmu)^\top \bSigma^{-1}(\x - \bmu)\right), \quad \x\in\mathbb{R}^p,
\end{equation}
where $g(t)$ is a non-negative Lebesgue measurable function on $[0,\infty)$ such that $\int_0^\infty t^{p/2-1} g(t) dt < \infty$ and $|\bSigma|$ denotes the determinant of matrix $\bSigma$.  Moreover, $$c_p = \frac{\Gamma(p/2)}{\pi^{p/2}} \left( \int_0^\infty t^{p/2-1} g(t) dt \right)^{-1}$$ is the normalizing constant, with $\Gamma(\cdot)$ representing the complete gamma function. We will use the notation $\X\sim E\ell_p(\bmu,\bSigma; g)$.

Members of the elliptical family of distributions are characterized by their density generating function $g$. Some examples of the elliptical family of distributions are: 
\begin{itemize}[leftmargin=.4cm]
\item The {\it multivariate normal} distribution, $\X\sim\N_p(\bmu,\bSigma)$, with mean $\bmu$ and variance-covariance matrix $\bSigma$, arises when the dgf takes the form $g(t)=\exp(-t/2), t \geq 0$.
\item The {\it multivariate Student-t} distribution, $\X\sim t_p(\bmu,\bSigma,\nu)$, where $\bmu$ is the location parameter, $\bSigma$ is the scale matrix, and $\nu$ is called the degrees of freedom, is obtained when $g(t)= (1+t/\nu)^{-(\nu+p)/2}, t\geq 0$.
\item The {\it multivariate power exponential}, $\X\sim \mbox{PE}_p(\bmu, \bSigma, \beta)$, with kurtosis parameter $\beta > 0$. In this case, $g(t)= \exp(-t^\beta/2), t\geq 0$. A particular case of the power exponential distribution is the normal distribution, which arises when $\beta=1$. 
\item The {\it multivariate slash}, $\X \sim \mbox{SL}_p(\bmu, \bSigma, \nu)$, we get a random variable with multivariate slash distribution when $g(t)=\int_0^1 u^{\nu+p/2-1} \exp\{-ut/2\} du, t\geq 0, \nu>0$.
\item The {\it multivariate Pearson VII} distribution, $\X\sim\mbox{PVII}_p(\bmu,\bSigma,m,\nu)$, with parameters $\bmu\in\mathbb{R}^p$, $\bSigma\in\mathbb{R}^{p\times p}$, $m>p/2$, and $\nu>0$ is obtained when $g(t)=(1+t/\nu)^{-m}, t \geq 0$.
\end{itemize}
For more distributions belonging to this family, please
see \cite{fang2018symmetric}.

\subsection{Truncated Elliptical Family of Distributions}
Let ${A}\subseteq \mathbb{R}^p$ be a measurable set. We say that a random vector $\Y\in\mathbb{R}^p$ has truncated elliptical distribution with support ${A}$, location parameter $\bmu\in\mathbb{R}^p$, scale parameter $\bSigma\in\mathbb{R}^{p\times p}$ and dgf $g$, if its pdf is given by
\begin{equation}\label{truncated}
f_\Y(\y) = \frac{g\left((\y-\bmu)^\top\bSigma^{-1}(\y-\bmu)\right)}{\int_{A} g\left((\y-\bmu)^\top\bSigma^{-1}(\y-\bmu)\right) d\y} = \frac{f_\X(\y)}{\mbox{Pr}(\X\in{A})}, \quad \y\in {A},
\end{equation}
where $\X \sim \mbox{E}\ell_p(\bmu, \bSigma; g)$. We use the notation $\Y\sim \mbox{TE}\ell(\bmu,\bSigma; g, A)$. Notice that the pdf of $\Y$ is written as the ratio between the pdf of $\X\sim \mbox{E}\ell_p(\bmu,\bSigma; g)$ and $\mbox{Pr}(\X\in {A})$, so the pdf of $\Y$ exists if the pdf of $\X$ does, which occurs if $\bSigma$ is a positive-definite \citep[see, ][]{moran2019new}. The variable $\Y$ is also said to be an elliptical distribution truncated on $A$, being represented by $\Y = \X |\, (\X \in A)$.

As in the elliptical family of distributions, the dgf $g$ determines any distribution within the truncated elliptical class of distributions, for example, if $g(t)= (1+t/\nu)^{-(\nu+p)/2}, t\geq 0, \nu >0$, then $\Y$ has TMVT distribution. We will denote the different members of the truncated elliptical family defined in the subsection before as $\Y \sim \mbox{TN}_p(\bmu,\bSigma; A)$ for the TMVN distribution, $\Y\sim \mbox{T}t_p(\bmu,\bSigma,\nu;  A)$ for the TMVT distribution, $\Y\sim \mbox{TPE}_p(\bmu, \bSigma, \beta; A)$ for the truncated multivariate power exponential, $\Y \sim \mbox{TSL}_p(\bmu, \bSigma, \nu; A)$ for the truncated multivariate slash distribution, and $\Y\sim\mbox{TPVII}_p(\bmu, \bSigma, m, \nu; A)$ for the truncated multivariate Pearson VII distribution. 


\subsection{Slice Sampling Algorithm}

Introduced by \cite{neal2003slice}, the slice sampling algorithm is a Markov Chain Monte Carlo (MCMC) method for drawing random samples from a given distribution. The algorithm's idea is to sample uniformly from the $(p+1)$-dimensional region under the graph of $f(\x)$, a non-negative function proportional to the pdf of $\X$. Hence, let $Y$ be an auxiliary variable such that the joint pdf of $\X$ and $Y$ is uniform over the region $U=\{(\x,y): 0 < y < f(\x) \}$, i.e., $f_{X,Y}(\x,y)\propto \mathbb{I}\left(0 < y < f(\x)\right)$, with $\mathbb{I}(\cdot)$ being the indicator function. Therefore, we can obtain samples from the distribution of $\X$ by sampling jointly $(\x,y)$ and then ignoring $y$ values.

\begin{figure}[ht]
\caption{Slice sampling algorithm for univariate random variables.}\label{slice}
\centering
\includegraphics[scale=.7]{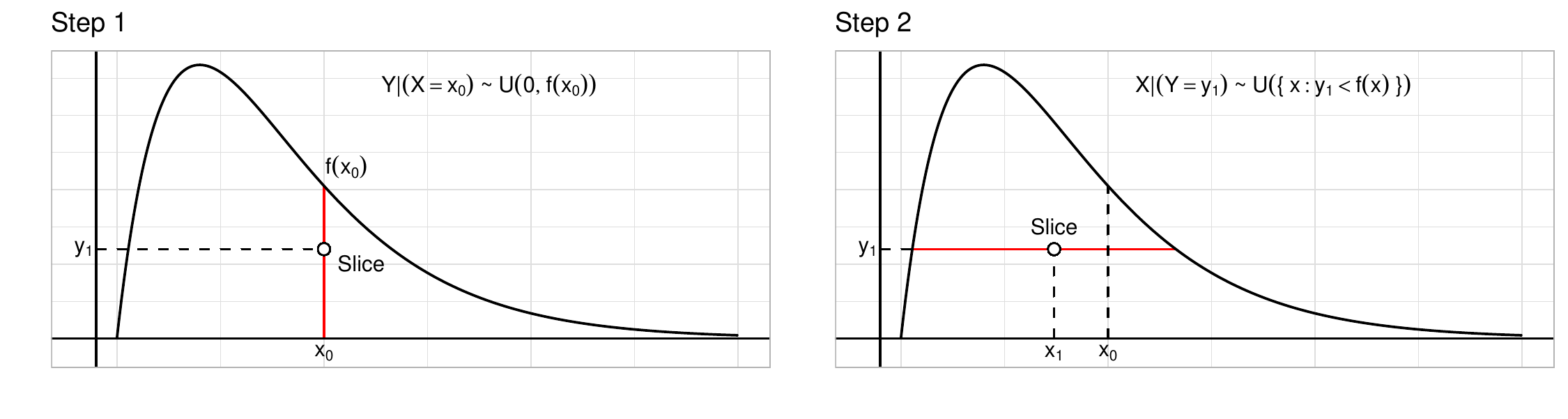}
\end{figure}

Note that generating independent random points uniformly distributed on $U$ may not be easy. To overcome this problem, \cite{neal2003slice} defined a Markov Chain that converges to an uniform distribution, in the same manner than the Gibbs sampling or Metropolis-Hastings algorithms. Then, considering Gibbs sampler steps, the slice sampling algorithm at iteration $k$ works as follows: given the current value of $\x_{k-1}$ sample $y_k$ from $Y|(\X=\x_{k-1}) \sim \mathcal{U}\left(0, f(\x_{k-1})\right)$, then draw $\x_k$ from the conditional distribution of $\X$ given $y_k$, which is uniform over the region $S_k=\{\x: y_k < f(\x)\}$, i.e., $\X|(Y=y_k) \sim \mathcal{U}(\{\x: y_k < f(\x)\})$, for all $k=1, 2, \ldots,n$, where $n$ is the desired sample size.

Figure \ref{slice} shows the steps of the slice sampling algorithm for $X$ being a univariate random variable. Given an initial value $X=x_0$, we draw $y_1$ uniformly over the interval $\left(0,f(x_0)\right)$ and then we sample $x_1$ from the conditional distribution of $X|(Y=y_1)$, i.e., uniformly over the interval $S_1=\{x: y_1<f(x)\}$. These two steps are repeated $n$ times.


\section{Sampling from the Truncated Elliptical Family of Distributions}\label{sliceSampling}

Next, we describe the proposed slice sampling algorithm with Gibbs sampler steps to generate samples
from a multivariate elliptical distribution with strictly decreasing dgf. Without loss of generality, we first consider a $p$-variate truncated elliptical distribution with zero location parameter, positive-definite scale matrix $\mathbf{R} \in \mathbb{R}^{p\times p}$, dgf $g$, and truncation region ${A}=\{\x: \mathbf{a}< \x < \mathbf{b}\}$, $\mathbf{a} < \mathbf{b}\in\mathbb{R}^p$, in other words, we will consider $\X\sim \mbox{TE}\ell_{p}(\mathbf{0}, \mathbf{R}; g, A)$. Here $\mathbf{R}$ is a correlation matrix, such that the scale matrix can be written as  $\bSigma = \bLambda \mathbf{R} \bLambda$, where $\bLambda = \mathrm{diag}(\sigma_{11},\ldots,\sigma_{pp})$. The pdf of $\X$ is given by
\begin{equation}
f_\X(\x) \propto g\left(\x^\top\textbf{R}^{-1}\x\right)\mathbb{I}\left(\x\in {A}\right),
\end{equation}

Now, in order to sample uniformly from the $(p + 1)$-dimensional region under the plot of $f_\X(\x)$, we introduce an auxiliary variable $Y$, such that the joint pdf of $\X$ and $Y$ is
\begin{equation}\label{xyjoint}
f_{\X,Y}(\x,y) \propto \mathbb{I}\left(0 < y < g\left(\x^\top\textbf{R}^{-1}\x\right) \right)\mathbb{I}\left(\textbf{a} < \x < \textbf{b}\right).
\end{equation}

It is enough to calculate the conditional distributions of $Y |\X$ and $\X|Y$ in order to established our slice
sampling algorithm with Gibbs steps to generate independent random observations from the pdf in (\ref{xyjoint}). These are given by:
\begin{eqnarray*}
f_{Y|\X}(y|\x) &\propto& \mathbb{I}\left(0 < y < g\left(\x^\top\textbf{R}^{-1}\x\right) \right) \quad \mbox{ and } \quad\\ f_{\X|Y}(\x|y) &\propto& \mathbb{I}\left(\big\{\x : y < g\left(\x^\top\textbf{R}^{-1}\x\right)\big\}\cap \{\textbf{a} < \x < \textbf{b}\} \right).
\end{eqnarray*}

Note that sampling $y$ from the distribution of $Y|(\X=\x)$ is straightforward, but sampling from $\X|(Y=y)$ is not trivial. Thus, we use the idea of \cite{ho2012some}, that consists in sampling each element of $\X$ given the remaining elements, i.e., sampling $X_j$ given $\x_{-j} = (x_1, \ldots, x_{j-1}, x_{j+1}, \ldots, x_p)^\top$ and $y$, for all $j=1,\ldots, p$. Hence, the following steps are performed to draw a random number from the distribution of $X_j|\x_{-j},y$.
\begin{enumerate}[leftmargin=0.6cm]
\item Let $\kappa_y = g^{-1}(y)$. Since $g$ is a strictly decreasing function, it follows that $y < g(\x^\top\textbf{R}^{-1}\x)$ is equivalent to $\kappa_y > \x^\top\textbf{R}^{-1}\x$.

\item Write $\x^\top\textbf{R}^{-1}\x = \rho^{jj} \left(x_j - \lambda_j \right)^2 - \rho^{jj}\lambda_j^2 + \eta_j$, where $\rho^{ij}$ is the $(i,j)$th element of the inverse of $\textbf{R}$, $\eta_j = \sum_{t\neq j}\sum_{r\neq j} x_t x_r\rho^{tr}$ and $\lambda_j = - \frac{1}{\rho^{jj}}\sum_{r\neq j} x_r\rho^{jr}$.

\item Combining items 1 and 2, we obtain that $\lambda_j-\tau_j < x_j < \lambda_j + \tau_j$, where $\tau_j = \left( \lambda_j^2 + \frac{1}{\rho^{jj}}\left(\kappa_y - \eta_j \right)\right)^{1/2}$

\item Because $x_j\in(a_j,b_j)$, thereby $a_j^*=\max(a_j, \lambda_j-\tau_j) < x_j < \min(b_j, \lambda_j+\tau_j)=b_j^*$.
\end{enumerate}

Therefore, the steps to draw $n$ samples from a $p$-variate truncated elliptical distribution $\X\sim \mbox{TE}\ell_{p}(\textbf{0}, \textbf{R};g, A)$ are summarized in Algorithm 1. As seen, only univariate uniform simulations are involved in the algorithm which are fast to compute. Note also that the assumption that the dgf $g$ is strictly decreasing has been used in step 1. A general case can be easily considered by studying the extrema points of $g$.

\begin{algorithm}[ht]
 \caption{Slice sampling algorithm}\label{alg1}
\SetAlgoLined
\KwIn{Sample size $n\geq 1$, initial value $\x_0 \in\mathbb{R}^p$, scale matrix $\textbf{R}\in\mathbb{R}^{p\times p}$, lower bound $\textbf{a}\in\mathbb{R}^p$, upper bound $\textbf{b}\in\mathbb{R}^p$ and strictly decreasing dgf $g(t), t\geq 0$.}
 Initialization\;
 \For{$i \gets 1$ \textbf{to} $n$}{
 Sample $y_i$ from  $Y|\x_{i-1} \sim\mathcal{U}(0, g(\x_{i-1}^\top\textbf{R}^{-1}\x_{i-1}))$\;
 $\kappa_y \gets g^{-1}(y_i)$\;
 
 \For{$j \gets 1$ \textbf{to} $p$}{
  $\eta_j \gets \displaystyle \sum_{t\neq j}\sum_{r\neq j} x_t x_r\rho^{tr}; \quad$ 
  $\lambda_j \gets \displaystyle -\frac{1}{\rho^{jj}}\sum_{r\neq j} x_r\rho^{jr}; \quad$
  $\tau_j \gets \displaystyle \left( \lambda_j^2 + \frac{1}{\rho^{jj}}\left(\kappa_y - \eta_j \right)\right)^{1/2}$\;
  Draw $x_j$ from $X_j|\x_{-j},y\sim \mathcal{U}(\max(a_j, \lambda_j-\tau_j), \min(b_j, \lambda_j+\tau_j))$\;
  $\x_i[j] \gets x_j; \quad$   
  $\X[i,j] \gets x_j$\;
  }
}
\KwResult{$\X$}
\end{algorithm}

Moreover, members of the truncated elliptical family of distributions are closed under affine transformations \citep{fang2018symmetric}. Hence drawing samples from $\Y\sim\mbox{TE}\ell_{p}(\bmu,\bSigma; g, (\textbf{a},\textbf{b}))$ may be readily done by sampling first from $\X\sim\mbox{TE}\ell_{p}(\textbf{0},\textbf{R}; g, (\textbf{a}^*,\textbf{b}^*))$, and then recovery $\Y$ by the following transformation $\Y = \bmu + \bLambda \X$, such that $\bSigma=\bLambda\textbf{R}\bLambda$, $\textbf{a}^*= \bLambda^{-1}\left(\textbf{a} - \bmu\right)$, and $\textbf{b}^*= \bLambda^{-1}\left(\textbf{b} - \bmu\right)$.

\subsection{ \textsf{R} function and Examples}

Algorithm \ref{alg1} and the transformation described previously were implemented in the \textsf{R} package \textbf{relliptical}. Its main function for random number generation is called \texttt{rtelliptical}, whose signature is the following.

\begin{lstlisting}[language=R]
rtelliptical(n=1e4, mu=rep(0,length(lower)), Sigma=diag(length(lower)), lower,
             upper=rep(Inf,length(lower)), dist="Normal", nu=NULL, expr=NULL, 
             gFun=NULL, ginvFun=NULL, burn.in=0, thinning=1)
\end{lstlisting}

In this function, $n \geq 1$ is the number of observations to be sampled, \texttt{nu} is the additional parameter or vector of parameters depending on the distribution of $\X$, \texttt{mu} is the location parameter, \texttt{Sigma} is the positive-definite scale matrix, and \texttt{lower} and \texttt{upper} are the lower and upper truncation points, respectively. The truncated normal, Student-$t$, power exponential, Pearson VII, slash, and contaminated normal distributions can be specified through the argument \texttt{dist}. 

The following examples illustrate the function \texttt{rtelliptical}, for drawing samples from truncated bivariate distributions with location parameter $\bmu=(0,0)^\top$, scale matrix elements $\sigma_{11} = \sigma_{22} = 1$, and $\sigma_{12} = \sigma_{21} = 0.70$, and truncation region ${A}=\{\x: \textbf{a}< \x < \textbf{b}\}$, with $\textbf{a}=(-2,-2)^\top$ and $\textbf{b}=(3,2)^\top$. The distributions considered are the predefined ones in the package.

\begin{itemize}[leftmargin=.4cm]
\item Truncated normal
\begin{lstlisting}[language=R]
rtelliptical(n=1e4, mu=c(0,0), Sigma=matrix(c(1,0.7,0.7,1),2,2), lower=c(-2,-2), 
             upper=c(3,2), dist="Normal")
\end{lstlisting}

\item Truncated Student-$t$ with $\nu=3$ degrees of freedom
\begin{lstlisting}[language=R]
rtelliptical(n=1e4, mu=c(0,0), Sigma=matrix(c(1,0.7,0.7,1),2,2), lower=c(-2,-2), 
             upper=c(3,2), dist="t", nu=3)
\end{lstlisting}

\item Truncated power exponential with kurtosis $\beta=2$
\begin{lstlisting}[language=R]
rtelliptical(n=1e4, mu=c(0,0), Sigma=matrix(c(1,0.7,0.7,1),2,2), lower=c(-2,-2), 
             upper=c(3,2), dist="PE", nu=2)
\end{lstlisting}

\item Truncated Pearson VII with parameters $m=5/2$ and $\nu=3$
\begin{lstlisting}[language=R]
rtelliptical(n=1e4, mu=c(0,0), Sigma=matrix(c(1,0.7,0.7,1),2,2), lower=c(-2,-2), 
             upper=c(3,2), dist="PVII", nu=c(2.50, 3.0))
\end{lstlisting}

\item Truncated slash with 3/2 degrees of freedom
\begin{lstlisting}[language=R]
rtelliptical(n=1e4, mu=c(0,0), Sigma=matrix(c(1,0.7,0.7,1),2,2), lower=c(-2,-2), 
             upper=c(3,2), dist="Slash", nu=1.50)
\end{lstlisting}

\item Truncated contaminated normal with $\nu = 0.70$ and $\rho = 0.20$
\begin{lstlisting}[language=R]
rtelliptical(n=1e4, mu=c(0,0), Sigma=matrix(c(1,0.7,0.7,1),2,2), lower=c(-2,-2), 
             upper=c(3,2), dist="CN", nu=c(0.70, 0.20))
\end{lstlisting}
\end{itemize}

Note that, no additional arguments are passed for the TMVN distribution. In the opposite way, for the
truncated contaminated normal and Pearson VII distributions, \texttt{nu} is a vector of length two, and for the remaining distributions, this parameter is a non-negative scalar. An important remark is that exists closed form expressions to compute $\kappa_y = g^{-1}(y)$ for the normal, Student-$t$, power exponential, and Pearson VII distributions, however, the contaminated normal and slash distributions require numerical methods for this purpose. This value is calculated as the root of the function $g(t) - y = 0, t \geq 0$, through the Newton-Raphson algorithm for the contaminated normal, and using Brent’s method \citep{brent2013algorithms}, for the slash distribution, a mixture of linear interpolation, inverse quadratic interpolation, and the bisection method.

This function also allows generating random numbers from other truncated elliptical distributions not
specified in the \texttt{dist} argument, by supplying the dgf through arguments either \texttt{expr} or \texttt{gFun}. The easiest way is to provide the dgf expression to argument \texttt{expr} as a character. The notation used in \texttt{expr} needs to be understood by package \textbf{Ryacas0} \citep{andersen2020ryacas}, and the  \textsf{R} environment. For instance, for the dgf $g(t) = e^{-t}$, the user must provide \verb'expr = "exp(1)^(-t)"'. For this case, when a character expression is provided to \texttt{expr}, the algorithm tries to compute a closed-form expression for the inverse function of $g(t)$, however, this is not always possible (a warning message is returned). On the other hand, if it is no possible to pass an expression to \texttt{expr}, due to the complexity of the expression, the user may provide a custom  \textsf{R} function to the \texttt{gFun} argument. By default, its inverse function is approximated numerically, however, the user may also provide its inverse to the \texttt{ginvFun} argument to gain some computational time. When \texttt{gFun} is provided, arguments dist and \texttt{expr} are ignored.

For example, to generate samples from the bivariate truncated logistic distribution with same parameters as
before, and which has dgf $g(t)=e^{-t}/(1+e^{-t})^2, t\geq 0$, we can run the following code.

\begin{lstlisting}[language=R]
rtelliptical(n=1e4, mu=c(0,0), Sigma=matrix(c(1,0.7,0.7,1),2,2), lower=c(-2,-2), 
             upper=c(3,2), expr="exp(1)^(-t)/(1+exp(1)^(-t))^2")
\end{lstlisting}

Another distribution that belongs to the elliptical family is the Kotz-type distribution with parameters $r>0, s>0$, and $2N + p > 2$, whose dgf is $g(t)=t^{N-1} e^{-r t^s}, t\geq 0$ \citep{fang2018symmetric}.  For this distribution, $g(t)$ is not strictly decreasing, however, for $(2-p)/2 < N \leq 1$, it holds. Hence, our proposal works for $r > 0$, $s > 0$, and $(2 - p)/2 < N \leq 1$. For this type of more complex dgf, it is advisable to pass it through the \texttt{gFun} argument as an  \textsf{R} function (with other parameters as fixed values). In the following example, we draw
samples from a bivariate Kotz-type distribution with settings as before, and extra parameters $r = 2, s = 1/4$, and $N = 1/2$.

\begin{lstlisting}[language=R]
rtelliptical(n=1e4, mu=c(0,0), Sigma=matrix(c(1,0.7,0.7,1),2,2), lower=c(-2,-2), 
             upper=c(3,2), gFun=function(t){t^(-1/2)*exp(-2*t^(1/4))})
\end{lstlisting}

\begin{figure}[ht]
    \centering
    \caption{Scatterplot and marginal histograms for the $n = 10^4$ observations sampled for some bivariate truncated elliptical distributions.}
    \includegraphics[scale=.73]{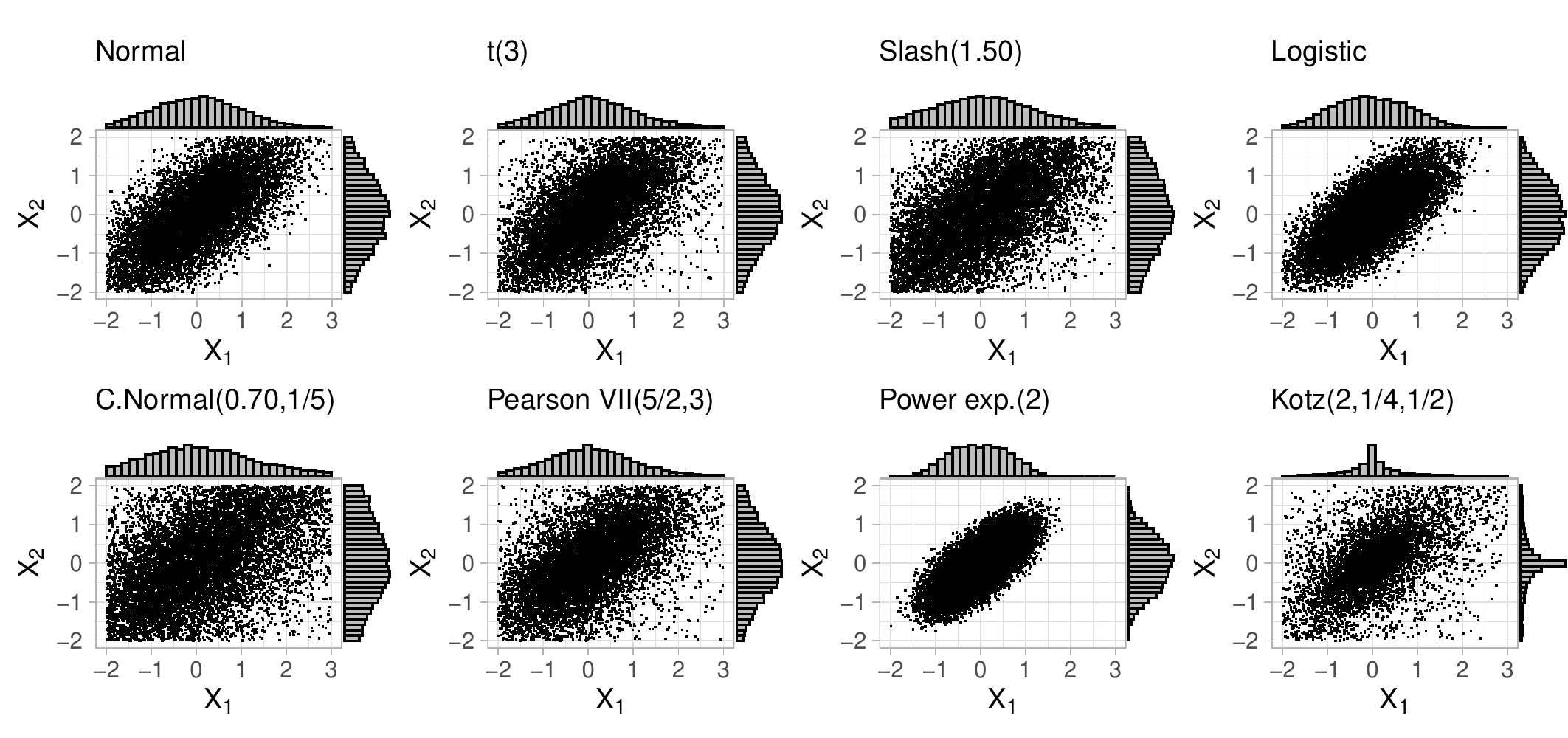}
    \label{fig_rng}
\end{figure}

Figure \ref{fig_rng} shows the scatterplot and marginal histograms for the $n = 10^4$ observations sampled from each of the truncated bivariate distributions referred above. 

As mentioned by \cite{robert2010introducing} and \cite{ho2012some}, the slice sampling algorithm with Gibbs steps generates random samples conditioned on previous values, resulting in a sequence of correlated samples. Thus, it is essential to analyze the dependence effect of the proposed algorithm. Figure \ref{fig_acf} in Section \ref{AppA} displays the autocorrelation plots for each one of the distributions, where we notice that the autocorrelation drops quickly and becomes negligibly small when lags become large, evidencing well mixing and quickly converging for these examples. If necessary, initial observations can be discarded by means of \texttt{the burn.in} argument. Finally, autocorrelation can be decimated by setting the \texttt{thinning} argument. Thinning consists in picking separated points from the sample, at each $k$th step. The thinning factor reduces the autocorrelation of the random points in the Gibbs sampling process. As natural, this value must be an integer greater than or equal to 1.


\section{Moments of Truncated Multivariate Elliptical Distributions}\label{moments}

This section describes an algorithm to compute the first two moments and the variance-covariance matrix of a random vector, whose distribution belongs to the elliptical family. Furthermore, we are going to apply this algorithm to some well-known distributions. Let $\X$ be a $p$-variate random vector that follows a truncated multivariate elliptical distribution with location parameter $\bmu\in\mathbb{R}^p$, positive-definite scale matrix $\bSigma\in\mathbb{R}^{p\times p}$, dgf $g$, and support ${A}\subseteq \mathbb{R}^p$, i.e., $\X\sim\mbox{TE}\ell_p(\bmu,\bSigma; g, A)$. The more straightforward approach for this problem is to use Monte Carlo integration. Following this approach, the estimates are given by
\begin{equation}
\widehat{\mathbb{E}(\X)} = \frac{1}{n}\sum_{i=1}^n \x_i, \quad \widehat{\mathbb{E}(\X\X^\top)}=\frac{1}{n}\sum_{i=1}^n \x_i\x_i^\top , \quad \widehat{\mathrm{Cov}(\X)} = \widehat{\mathbb{E}(\X\X^\top)} - \widehat{\mathbb{E}(\X)}\widehat{\mathbb{E}(\X)}^\top,
\end{equation}
where $\x_i$ is the $i$th sample of the random vector $\X$ draws from $\mbox{TE}\ell_p(\bmu,\bSigma; g, A)$. However, it is well-known that the execution time needed to perform Monte Carlo integration depends on the algorithm employed to draw samples, the number of random points ($n$) used in the approximation, and the length of the random vector ($p$). Then, it depends on some variables that might represent a considerable computational effort. Nevertheless, we can save time when the random vector $\X$ has non-truncated components following the idea of \cite{galarza2020moments}. They proposed to decompose $\X$ into two vectors, $\X_1$ and $\X_2$, in such a way that $\X_1$ is the random vector of truncated variables and $\X_2$ is the non-truncated part, and then compute the moments for the truncated variables using any method and the remaining moments using properties of the conditional expectation. Before showing our algorithm, we state an extremely important result.

\begin{proposition}[\textit{Marginal and conditional distribution of the Elliptical family}] \label{remark1}
Let $\X\in\mathbb{R}^p$ be partitioned into two vectors, $\X_1\in\mathbb{R}^{p_1}$ and $\X_2\in\mathbb{R}^{p_2}$, such that $p=p_1+p_2$ and $\X=(\X_1^\top,\X_2^\top)^\top$ has joint multivariate elliptical distribution as follows 
\begin{equation*}
\X = \left(\begin{array}{c} \X_1 \\ \X_2 \end{array}\right) \sim \mbox{E}\ell_{p_1+p_2} \left(\bmu=\left(\begin{array}{c} \bmu_1 \\ \bmu_2 \end{array}\right), \bSigma = \left(\begin{array}{cc} \bSigma_{11} & \bSigma_{12} \\ \bSigma_{21} & \bSigma_{22}\end{array}\right); g^{(p_1+p_2)} \right),
\end{equation*}
where $\bmu_1\in\mathbb{R}^{p_1}$, $\bmu_2\in\mathbb{R}^{p_2}$ are location vectors, $\bSigma_{11} \in \mathbb{R}^{p_1\times p_1}, \bSigma_{22} \in \mathbb{R}^{p_2 \times p _2}, \bSigma_{12} \in \mathbb{R}^{p_1 \times p_2}, \bSigma_{21} \in \mathbb{R}^{p_2 \times p_1}$ are dispersion matrices, and $g^{(p_1+p_2)}$ is the dgf. \cite{fang2018symmetric} demonstrated that the elliptical family of distributions is closed under marginalization and conditioning. Hence, the distribution of $\X_1$ and $\X_2|(\X_1=\x)$ are also elliptical, with \\
$\X_1\sim \mbox{E}\ell_{p_1}\left(\bmu_{1}, \bSigma_{11}; g_1^{(p_1)}\right),$ \\
$\X_2|(\X_1=\x) \sim \mbox{E}\ell_{p_2} \left( \bmu_{2}+\bSigma_{21}\bSigma_{11}^{-1}(\x-\bmu_1), \bSigma_{22} - \bSigma_{21}\bSigma_{11}^{-1}\bSigma_{12}; g^{(p_2)}_{\x} \right).$
\end{proposition}

Therefore, considering that $\X_1$ is the vector of truncated variables with truncation region ${A}_1$ and $\X_2$ is the vector of non-truncated variables, by Proposition \ref{remark1} we have that 
\begin{eqnarray*}
&\X_1\sim \mbox{TE}\ell_{p_1}(\bmu_1, \bSigma_{11}; g_1^{(p_1)}, A_1) \quad \mbox{and} \\
&\X_2|(\X_1=\x) \sim \mbox{E}\ell_{p_2} \left( \bmu_{2}+\bSigma_{21}\bSigma_{11}^{-1}(\x-\bmu_1), \bSigma_{22} - \bSigma_{21}\bSigma_{11}^{-1}\bSigma_{12}; g^{(p_2)}_{\x} \right).
\end{eqnarray*}

Let $\bxi_1=\mathbb{E}(\X_1|\X_1\in {A}_1)$ and $\bOmega_{11}=\mathrm{Cov}(\X_1|\X_1\in {A}_1)$. Then, it follows that $\mathbb{E}(\X|\X\in {A}) = \mathbb{E}(\mathbb{E}(\X|\X_1)|\X_1\in {A}_1)$, that is
\begin{eqnarray}
\E(\X|\X\in {A}) = \mathbb{E}\left(\begin{array}{c} \X_1 \\ \bmu_{2} + \bSigma_{21}\bSigma_{11}^{-1}(\X_1 - \bmu_1)
\end{array}\Big| \X_1\in {A}_1 \right) = \left(\begin{array}{c} \bxi_1 \\ \bmu_2 + \bSigma_{21}\bSigma_{11}^{-1}(\bxi_1 - \bmu_1)  \end{array}\right). \label{mean}
\end{eqnarray}

On the other hand, we have that $\mathrm{Cov}(\X | \X\in {A}) = \mathrm{Cov}(\mathbb{E}(\X | \X_1) | \X_1 \in {A}_1) + \mathbb{E}(\mathrm{Cov}(\X | \X_1) | \X_1\in {A}_1)$, with
\begin{itemize}[leftmargin=0.4cm]
    \item $\mathrm{Cov}(\X_1,\mathbb{E}(\X_2|\X_1)|\X_1\in {A}_1) = \mathrm{Cov}(\X_1, \bSigma_{21}\bSigma_{11}^{-1} \X_1| \X_1 \in  {A}_1) = \bOmega_{11}\bSigma_{11}^{-1}\bSigma_{12},$
    \item $\mathrm{Cov}(\mathbb{E}(\X_2|\X_1)|\X_1\in {A}_1) = \mathrm{Cov}(\bSigma_{21}\bSigma_{11}^{-1}\X_1 | \X_1 \in {A}_1) = \bSigma_{21}\bSigma_{11}^{-1}\bOmega_{11}\bSigma_{11}^{-1}\bSigma_{12},$
    \item $\mathbb{E}(\mathrm{Cov}(\X_2|\X_1) | \X_1\in {A}_1) = \omega_{2.1}(\bSigma_{22} - \bSigma_{21} \bSigma_{11}^{-1} \bSigma_{12}),$
\end{itemize}
where $\omega_{2.1}=\E\left(h(\X_1)|\X_1\in {A}_1\right)$ is the expected value of a function $h$ of $\X_1$ depending on the conditional dgf $g_{\x_1}^{(p_2)}$. So, the variance-covariance matrix of $\X$ is given by
\begin{eqnarray}
\mathrm{Cov}(\X|\X\in {A}) = \left(\begin{array}{cc} \bOmega_{11} & \bOmega_{11}\bSigma_{11}^{-1}\bSigma_{12} \\ \bSigma_{21}\bSigma_{11}^{-1}\bOmega_{11} & \omega_{2.1}\bSigma_{22} - \bSigma_{21}\bSigma_{11}^{-1}\left(\omega_{2.1}\textbf{I}_{p_1} - \bOmega_{11}\bSigma_{11}^{-1}\right)\bSigma_{12}  \end{array}\right), \label{var}
\end{eqnarray}
Thereby, we just need Monte Carlo integration to approximate $\bxi_1$, $\bOmega_{11}$, and $\omega_{2.1}$ (if necessary). A brief summary of how our algorithm works is given in Algorithm \ref{alg2}.

\begin{algorithm}[ht]
 \caption{Mean and variance approximation}\label{alg2}
\SetAlgoLined
\KwIn{Sample size $n\geq 1$, location parameter $\bmu\in\mathbb{R}^p$, scale matrix $\bSigma\in\mathbb{R}^{p\times p}$, lower bound $\textbf{a}\in\mathbb{R}^p$, upper bound $\textbf{b}\in\mathbb{R}^p$ and dgf $g(t), t\geq 0$.}
 Identify: $\bmu_1, \bmu_2, \bSigma_{11}, \bSigma_{22}, \bSigma_{12}, {A}_1=\{\x_1: \textbf{a}_1<\x_1< \textbf{b}_1\}$\;
 Draw $\x_{1i}$ from $\X_1\sim \mbox{TE}\ell_{p_1}(\bmu_1, \bSigma_{11}; g^{p_1}, A_1), \quad i=1,\ldots, n$\; 
 $\widehat{\bxi}_1 \gets \displaystyle \frac{1}{n}\sum_{i=1}^n \x_{1i}; \quad$ 
 $\widehat{\bOmega}_{11} \gets \displaystyle \frac{1}{n}\sum_{i=1}^n \x_{1i}\x_{1i}^\top - \widehat{\bxi}_1\widehat{\bxi}_1^\top; \quad$
 $\widehat{\omega}_{2.1} \gets \displaystyle \frac{1}{n}\sum_{i=1}^n h(\x_{1i})$\;
 $\widehat{\E(\X)} \gets \left( \begin{array}{cc}
 \widehat{\bxi}_1 \\ \bmu_2 + \bSigma_{21}\bSigma_{11}^{-1}(\widehat{\bxi}_1 - \bmu_1) \end{array} \right)$\;
 $\widehat{\mathrm{Cov}(\X)} \gets \left(\begin{array}{cc} \widehat{\bOmega}_{11} & \widehat{\bOmega}_{11}\bSigma_{11}^{-1}\bSigma_{12} \\ \bSigma_{21}\bSigma_{11}^{-1}\widehat{\bOmega}_{11} & \widehat{\omega}_{2.1}\bSigma_{22} - \bSigma_{21}\bSigma_{11}^{-1}\left(\widehat{\omega}_{2.1}\textbf{I}_{p_1} - \widehat{\bOmega}_{11}\bSigma_{11}^{-1}\right)\bSigma_{12}  \end{array}\right)$\;
 $\widehat{\E(\X\X^\top)} \gets \widehat{\mathrm{Cov}(\X)} + \widehat{\E(\X)}\widehat{\E(\X)}^\top$\;
\KwResult{$\widehat{\E(\X)}, \widehat{\E(\X\X^\top)}, \widehat{\mathrm{Cov}(\X)}$}
\end{algorithm}

\subsection{Mean and Variance for the Truncated Elliptical Distributions} \label{meanvar}

Now, in this subsection, we analyze how Algorithm \ref{alg2} works for some specific distributions considering all the conditions used previously.

\begin{itemize}[leftmargin=0.4cm]
\item \textbf{Normal}: If $\X\sim\N_p(\bmu,\bSigma)$, the marginal distribution is $\X_1\sim \N_{p_1}(\bmu_1, \bSigma_{11})$ and the conditional distribution is $\X_2|(\X_1=\x) \sim \N_{p_2}(\bmu_{2.1}, \bSigma_{2.1})$, with $\bmu_{2.1} = \bmu_{2} + \bSigma_{21}\bSigma_{11}^{-1}(\x - \bmu_1)$ and $\bSigma_{2.1} = \bSigma_{22} - \bSigma_{21} \bSigma_{11}^{-1}\bSigma_{12}$. Then, With the above conditions, Algorithm \ref{alg2} firstly sample $\X_1$ from the truncated multivariate normal distribution with mean $\bmu_1$, covariance matrix $\bSigma_{11}$, truncation region ${A}_1$ and $\omega_{2.1}$ equal to 1.

\item \textbf{Student-$t$}: If $\X\sim t_p(\bmu, \bSigma, \nu), \nu>0$, the marginal and conditional distributions are $\X_1\sim t_{p_1}(\bmu_1,\bSigma_{11},\nu)$ and $\X_2|(\X_1=\x) \sim t_{p_2}(\bmu_{2.1}, \lambda\bSigma_{2.1}, \nu+p_1)$, respectively, such that $\bmu_{2.1}=\bmu_{2}+\bSigma_{21} \bSigma_{11}^{-1}(\x-\bmu_1)$, $\lambda=(\nu + \delta_1(\x))/(\nu + p_1)$, $\bSigma_{2.1} = \bSigma_{22} - \bSigma_{21}\bSigma_{11}^{-1}\bSigma_{12}$ and $\delta_1(\x) = (\x - \bmu_1)^\top \bSigma_{11}^{-1} (\x - \bmu_1)$. For this distribution $\E(\X)<\infty$, if $\nu>1$ and $\mathrm{Cov}(\X)<\infty$, if $\nu>2$. Therefore, the algorithm samples $\X_1$ from the truncated $t$ distribution with location parameter $\bmu_1$, scale matrix $\bSigma_{11}$, $\nu$ degrees of freedom, truncation region ${A}_1$, and $\omega_{2.1}$ computed by
$$\displaystyle\omega_{2.1}=\frac{\nu + \E(\delta_1(\X_1)|\X_1\in {A}_1)}{\nu + p_1 - 2}, $$ 
with $\E(\delta_1(\X_1)|\X_1\in {A}_1) = \mathrm{tr}(\bOmega_{11}\bSigma_{11}^{-1}) + (\bxi_1-\bmu_1)^\top\bSigma_{11}^{-1}(\bxi_1-\bmu_1)$. It is worth mention that for doubly truncated variables, the mean and the variance exist for all $\nu>0$. Then, if $\X$ has at least two doubly truncated variables, the mean and the variance-covariance matrix exist for all $\nu>0$. For more details about the existences of the moments see \cite{galarza2020moments}.

\item \textbf{Pearson VII}: If $\X\sim \mbox{PVII}_p(\bmu, \bSigma, m, \nu), m >p/2, \nu>0$, then $\E(\X)=\bmu$ and $\mathrm{Cov}(\X) = \frac{\nu}{2m-p-2} \bSigma$. In this case, $\E(\X)<\infty$, if $m>(p+1)/2$ and $\mathrm{Cov}(\X)<\infty$, if $m>(p+2)/2$. The marginal and the conditional distributions are $\X_1\sim \mbox{PVII}_{p_1}(\bmu_1,\bSigma_{11}, m-p_2/2, \nu)$ and $\X_2|(\X_1=\x) \sim \mbox{PVII}_{p_2}(\bmu_{2.1}, \bSigma_{2.1}, m, \nu+\delta_1(\x))$, respectively, such that $\bmu_{2.1} = \bmu_{2} + \bSigma_{21}\bSigma_{11}^{-1}(\x - \bmu_1)$, $\bSigma_{2.1} = \bSigma_{22} - \bSigma_{21} \bSigma_{11}^{-1}\bSigma_{12}$ and $\delta_1(\x) = (\x - \bmu_1)^\top \bSigma_{11}^{-1}(\x - \bmu_1)$. So, the proposed algorithm was implemented by sampling $\X_1$ from the truncated multivariate Pearson VII distribution with location parameter $\bmu_1$, scale matrix $\bSigma_{11}$, additional parameters $m-p_2/2 > p_1/2$, $\nu>0$, and truncation region ${A}_1$. The constant $\omega_{2.1}$ is
$$\omega_{2.1} = \frac{\nu + \E(\delta_1(\X_1)|\X_1\in {A}_1)}{2m - p_2 - 2},$$
where $\E(\delta_1(\X_1)|\X_1\in {A}_1)$ is given as in the Student-$t$ distribution. For this distribution, first and second moments for doubly truncated variables exist for all $m>p/2$. Then, if $\X$ has at least two doubly truncated variables, the mean and the variance exist for all $m>p/2$. For more details about the existence of the moments, see Appendix \ref{pvii}.

\item \textbf{Slash}: If $\X\sim \mbox{SL}_p(\bmu, \bSigma, \nu), \nu>0$, then $\mathbb{E}(\X) = \bmu$ and $\mathrm{Cov}(\X) = \frac{\nu}{\nu - 1}\bSigma$. In this case, $\mathrm{Cov}(\X) < \infty$, if $\nu > 1$. The marginal distribution is  $\X_1\sim \mbox{SL}_{p_1}(\bmu_1, \bSigma_{11}, \nu)$ and the conditional distribution is $\X_2|(\X_1=\x) \sim \mbox{E}\ell_{p_2}(\bmu_{2.1}, \bSigma_{2.1}; g^{(p_2)})$, such that $\bmu_{2.1}=\bmu_{2}+\bSigma_{21}\bSigma_{11}^{-1}(\x-\bmu_1)$, $\bSigma_{2.1}=\bSigma_{22} - \bSigma_{21} \bSigma_{11}^{-1}\bSigma_{12}$, $g^{(p_2)}(t) = \int_0^1 u^{\nu+p/2-1} \exp\{-u(t+\delta_1(\x))/2\}du$ and $\delta_1(\x) = (\x - \bmu_1)^\top \bSigma_{11}^{-1}(\x - \bmu_1)$. Note that $\X_2|\X_1$ does not follow slash distribution, but its distribution belongs to the elliptical family (see Appendix \ref{slash}). So, $\X_1$ is sampled from the truncated multivariate slash distribution with location parameter $\bmu_1$, scale matrix $\bSigma_{11}$, $\nu$ degrees of freedom and truncation region ${A}_1$. The constant $\omega_{2.1}$ is given by
$$\omega_{2.1} = \frac{\nu}{\nu-1} \E\left(\frac{\mbox{SL}_{p_1}(\X_1;\bmu_{1},\bSigma_{11},\nu-1)}{\mbox{SL}_{p_1}(\X_1;\bmu_{1},\bSigma_{11},\nu)}\Big|\X_1\in {A}_1 \right).$$
This constant can be also approximated via Monte Carlo integration. 

\item \textbf{Contaminated Normal}: If $\X\sim \mbox{CN}_p(\bmu, \bSigma, \nu, \rho), 0 <\nu,\rho <1$, then the distributions of $\X_1$ and $\X_2|(\X_1=\x)$ are $\mbox{CN}_{p_1}(\bmu_1,\bSigma_{11},\nu,\rho)$ and $\mbox{CN}_{p_2}(\bmu_{2.1},\bSigma_{2.1},\nu_{2.1},\rho)$, respectively, such that $\bmu_{2.1} = \bmu_{2} + \bSigma_{21}\bSigma_{11}^{-1}(\x-\bmu_1)$, $\bSigma_{2.1} = \bSigma_{22} - \bSigma_{21} \bSigma_{11}^{-1}\bSigma_{12}$, $\nu_{2.1} = \nu\phi_{p_1}(\x; \bmu_1, \rho^{-1} \bSigma_{11})/(\nu\phi_{p_1}(\x; \bmu_1, \rho^{-1} \bSigma_{11}) + (1-\nu)\phi_{p_1}(\x; \bmu_1, \bSigma_{11}))$ and $\phi_{p}(\x; \bmu, \bSigma)$ denotes the pdf of a $p$-variate normal distribution with mean $\bmu$, variance matrix $\bSigma$ evaluated at point $\x\in\mathbb{R}^{p_1}$. Thus, $\X_1$ is sampled from the truncated contaminated normal distribution with parameters $\bmu_1$, $\bSigma_{11}$, $\nu$ and $\rho$. The constant is  $\omega_{2.1}$ is $\omega_{2.1} = {\nu}_{2.1}^*/\rho + 1 - {\nu}_{2.1}^*$, where ${\nu}_{2.1}^*= \E(\nu_{2.1} | \X_1\in {A}_1)$, this value is also approximated via Monte Carlo integration.

\item \textbf{Power exponential}: If $\X\sim \mbox{PE}_p(\bmu,\bSigma,\beta), \beta>0$, then $\mathbb{E}(\X) = \bmu$ and $Cov(\X) = \omega \bSigma$, with $\omega = 2^{1/\beta} \Gamma(\frac{n+2}{2\beta})/ (n \Gamma(\frac{n}{2\beta}))$. The marginal distribution of $\X_1$ belongs to the elliptical family of distributions with dgf $g^{(p_1)}(t) = t^{\frac{p-p_1}{2}} \int_0^1 w^{\frac{p_1-p}{2}} (1-w)^{\frac{p-p_1}{2}-1} \exp\{-\frac{t^\beta}{2 w^\beta}\} dw$, $\X_1\sim\mbox{E}\ell_{p_1}(\bmu_1, \bSigma_{11}; g^{(p_1)})$, and the conditional distribution is $\X_2|(\X_1=\x) \sim \mbox{E}\ell_{p_2}(\bmu_{2.1}, \bSigma_{2.1}; g^{(p_2)})$ where $\bmu_{2.1} = \bmu_{2} + \bSigma_{21}\bSigma_{11}^{-1} (\x - \bmu_1)$, $\bSigma_{2.1} = \bSigma_{22} - \bSigma_{21} \bSigma_{11}^{-1}\bSigma_{12}$, $g^{(p_2)}(t) = \exp\{-\frac{1}{2}(t + \delta_1(\x))^\beta\}$ and $\delta_1(\x) = (\x-\bmu_1)^\top\bSigma_{11}^{-1} (\x-\bmu_1)$ is the squared Mahalanobis distance \citep{gomez1998multivariate}. Since sampling directly from the marginal distribution of $\X_1$ could be really complicated, we will use a different approach that consists of drawing points from the whole random vector of length $p$ and then approximate the moments using Monte Carlo integration. 
\end{itemize}

\subsection{\textsf{R} function and Examples}

The Algorithm \ref{alg2} for the distributions mentioned in subsection \ref{meanvar} has been implemented in the  \textsf{R} function \texttt{mvtelliptical}, whose signature together with default values is the following.

\begin{lstlisting}[language=R]
mvtelliptical(lower, upper=rep(Inf,length(lower)), mu=rep(0,length(lower)), 
              Sigma=diag(length(lower)), dist="Normal", nu=NULL, n=1e4, burn.in=0, 
              thinning=3)
\end{lstlisting}

The arguments \texttt{lower} and \texttt{upper} are the lower and upper truncation points of length $p$, respectively, \texttt{mu} is the location parameter of length $p$, \texttt{Sigma} is the $p\times p$ positive-definite scale matrix, \texttt{nu} is the additional parameter or vector of parameters depending on the dgf $g$. The argument \texttt{dist} indicates the distribution to be used. The parameters \texttt{n}, \texttt{burn.in}, and \texttt{thinning} are related to the Monte Carlo approximation, where \texttt{n} is the number of samples to be generated, \texttt{burn.in} is the number of samples to be discarded as burn-in phase, and \texttt{thinning} is a factor for reducing autocorrelation between observations.

\vspace*{0.2cm}
\noindent {\it Example 1}
\vspace*{0.2cm}

We illustrate how the method works considering a random vector of length 4 with truncated Student-$t$ distribution. In this example, the second variable is not truncated, and the others are doubly truncated. The objective is to study the performance of the estimates for the mean and the variance-covariance elements obtained through Algorithm \ref{alg2}, considering a different number of samples and thinning. After that, we compare those results with the estimates from the R functions \texttt{meanvarTMD} available in package \textbf{MomTrunc} and \texttt{TT.moment} from package \textbf{TTmoment}.

\begin{figure}[ht]
    \centering
        \caption{Boxplot based on 100 estimates of the truncated mean. The red dashed line represents the median of the estimates obtained from function \texttt{TT.moment}.}
    \includegraphics[scale=.68]{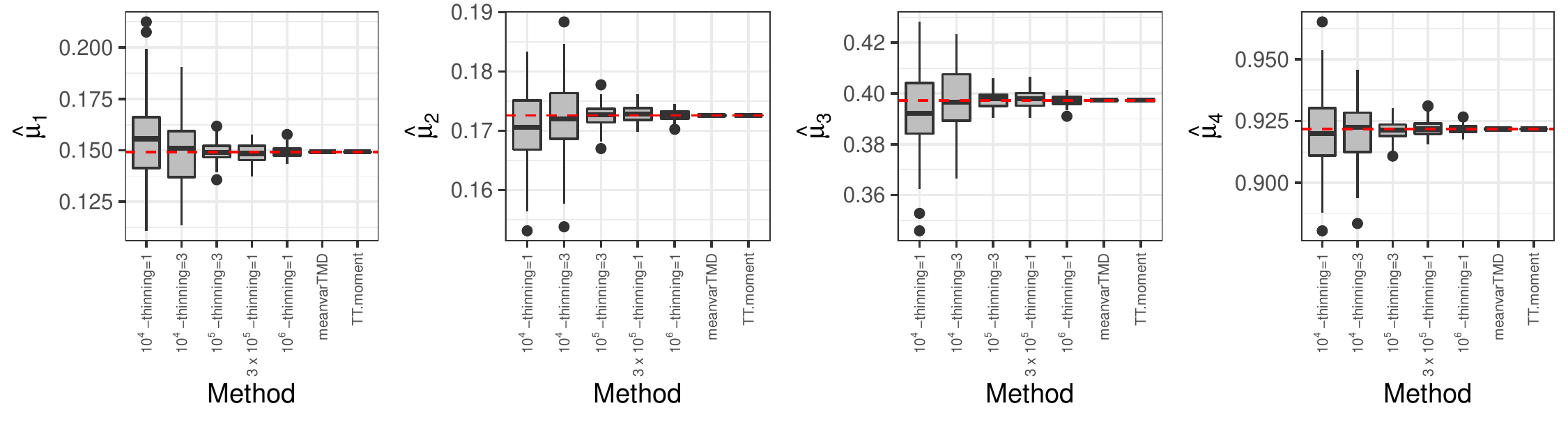}
    \label{figT_mean}
\end{figure}

Figure \ref{figT_mean} displays the boxplot for each element of the mean vector based on 100 estimates obtained through our proposal considering $n = 10^4$ with thinning=1 and =3, $n=10^5$ with thinning=3, $n=3\times 10^5$ with thinning=1, and $10^6$ samples with no thinning (=1). Also displays the results came from the function \texttt{meanvarTMD} and the function \texttt{TT.moment}. The red dashed line represents the median of the estimates achieved from the \texttt{TT.moment} function. It is possible to observe that for the case of $n=10^4$, the estimates obtained with no thinning have more variability than those with thinning=3 (observations with lower autocorrelation). The median of \texttt{TT.moment} estimates is closer to the median of our method in most cases, except for $n=10^4$ with thinning=1. As expected, the variability in the estimates was reduced when the sample size was increased. The distribution of the estimates from our algorithm with $10^5$ samples and thinning=3 was similar to the distribution considering $n = 3\times 10^5$ and no thinning. Recall that both methods needed to generate the same number of samples; the only difference here is that the first one (thinning=3) will need less memory space than the other one. The best results were obtained throughout \texttt{TT.moment} and \texttt{meanvarTMD} functions. Those results are comparable with the estimates achieved from our proposal with $n=10^6$ and no thinning. 

\begin{figure}[ht]
    \centering
        \caption{Boxplot based on 100 estimates of the variance-covariance elements. The red dashed line represents the median of the estimates obtained from function \texttt{TT.moment}.}
    \includegraphics[scale=.68]{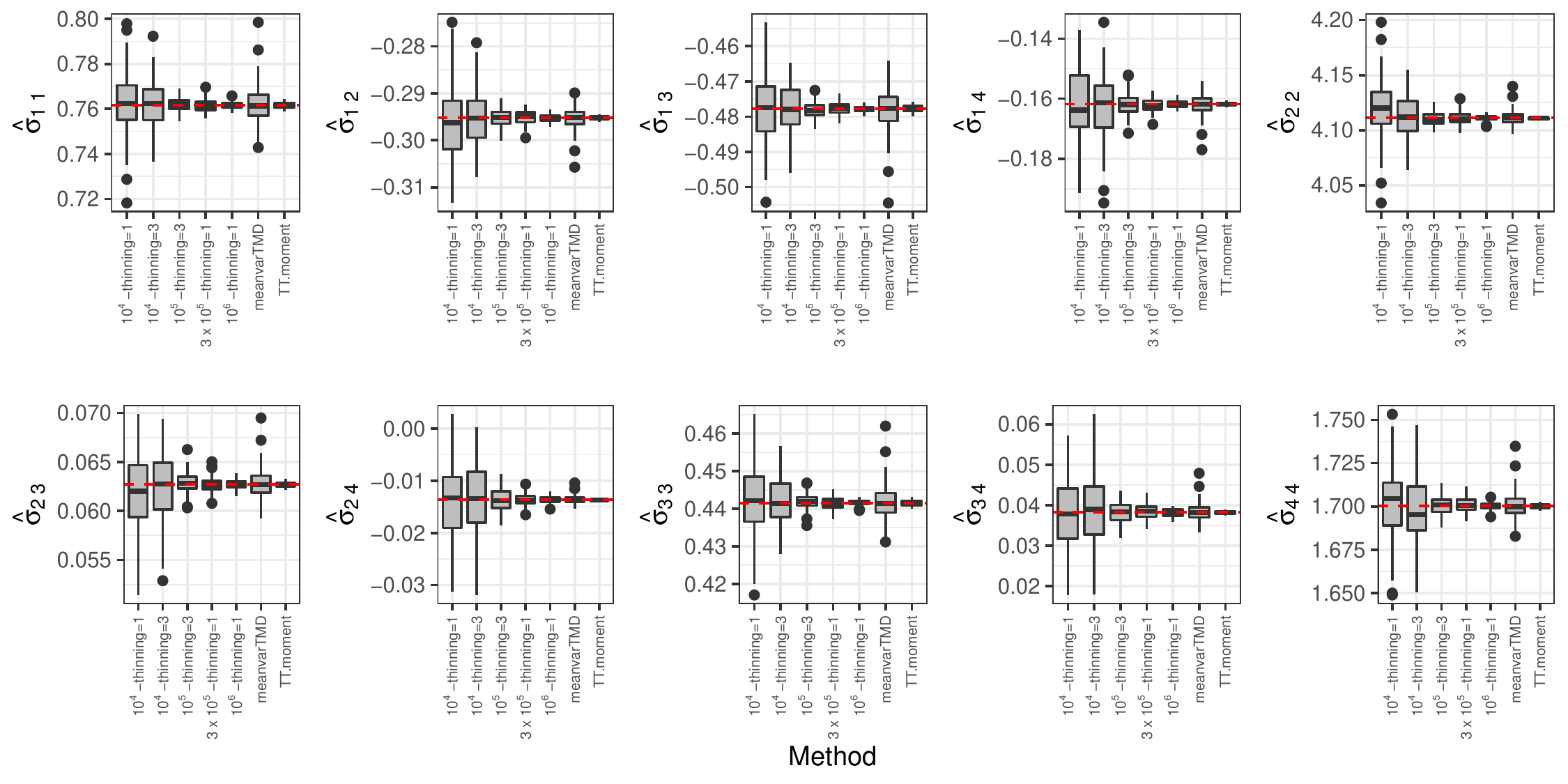}
    \label{figT_var}
\end{figure}

Figure \ref{figT_var} shows the boxplot for the variance-covariance elements of the truncated random vector considering each method described above. We noticed a slight reduction in the variability of the estimates when considering a thinning equal to 3. Another interesting fact is observed when we set $n=10^5$ and thinning=3; in this case, it returned similar results than estimate the covariances from MC with $3\times 10^5$ samples and no thinning. The estimates achieved through our proposal considering $n = 10^6$ are comparable with the results from \texttt{TT.moment}. The estimates obtained from \texttt{meanvarTMD} are similar to those from MC with $n=10^5$ and thinning=3 in most cases, except for $\sigma_{11}, \sigma_{33}$, and $\sigma_{13}$. For these parameters, our method showed better performance.

\vspace*{0.2cm}
\noindent {\it Example 2}
\vspace*{0.2cm}

In the previous example, it was observed that the estimates obtained from Algorithm \ref{alg2} with $n = 10^5$ and thinning=3 are good enough to estimate the mean and variance of a multivariate ($p=4$) variable with TMVT distribution, even though the best results were gotten through the \texttt{TT.moment} function. In this example, our goal is to analyze the execution time required for our method and the functions \texttt{meanvarTMD} and \texttt{TT.moment} to estimate the first two moments and the variance-covariance matrix of a $p$-variate random vector with TMVT distribution considering $p=50, 100, 150$. In each case, we set 10\%, 20\%, and 40\% of the variables doubly truncated. The methods were run in a Windows 10 machine using R 4.0.3 on an Intel Core i7-7700 Processor with 3.60 GHz, and 32 GB of RAM.

Table \ref{T-time} displays the median of the running time (in seconds) required for our algorithm and functions \texttt{meanvarTMD} and \texttt{TT.moment}. For our proposal were considered three scenarios $n=10^4$ with no thinning, $n=10^4$ with thinning=3, and $n=10^5$ with thinning=3. The results are based on 100 simulations, and they were computed through the \textsf{R} function \texttt{microbenchmark}. This table also shows the relative time computed, taking the time used by our method with $n = 10^5$ and thinning=3 as reference. We are going to refer to this configuration as the ``reference method". For our algorithm, we observed that the time required to estimate the moments depends only on the number of random observations sampled. Note that estimating the moments with $n = 10^4$ took 3.50\% of the time required for the reference method, and it is worth mention that the number of samples needed for the first method is 3.33\% the number of samples used for the reference one. Our proposal with $n = 10^4$ and thinning=3 already needed 10\% of the execution time used by the reference method. Observe that the only case where \texttt{meanvarTMD} was faster than the reference one is when the number of doubly truncated variables is equal to 5. It also seems that the time needed by the \texttt{meanvarTMD} function depends only on the number of doubly truncated variables. In all scenarios, the \texttt{TT.moment} function is much more time-consuming, e.g., for a random vector of length $p=100$ and 40 doubly truncated variables, it needed 28 times longer than the reference method. An additional example can be found in Appendix \ref{AppAx}. 

\begin{table}[ht]
\centering
\small
\caption{Median of the execution time (in seconds) based on 100 simulations.} \label{T-time}
\begin{tabular*}{\textwidth}{c@{\hspace{\tabcolsep}\extracolsep{\fill}}lccccccccc}
\toprule
\multirow{2}{*}{Method} & \multirow{2}{*}{Measure} & \multicolumn{3}{c}{$p=50$} & \multicolumn{3}{c}{$p=100$} & \multicolumn{3}{c}{$p=150$} \\
\cmidrule(lr){3-5}  \cmidrule(lr){6-8}  \cmidrule(l){9-11}
& & 10\% & 20\% & 40\% & 10\% & 20\% & 40\% & 10\% & 20\% & 40\% \\
\bottomrule
$n = 10^4$ & Median & 0.011 & 0.030 & 0.139 & 0.030 & 0.140 & 0.952 & 0.071 & 0.382 & 2.842 \\
thinning = 1 & R.Time & 0.035 & 0.035 & 0.034 & 0.036 & 0.035 & 0.034 & 0.036 & 0.034 & 0.034 \\
\midrule
$n = 10^4$ & Median & 0.031 & 0.084 & 0.404 & 0.085 & 0.405 & 2.820 & 0.199 & 1.118 & 8.461 \\
thinning = 3 & R.Time & 0.100 & 0.100 & 0.100 & 0.100 & 0.100 & 0.100 & 0.101 & 0.100 & 0.100 \\
\midrule
$n=10^5$ & Median & 0.314 & 0.844 & 4.042 & 0.846 & 4.044 & 28.217 & 1.974 & 11.182 & 84.619 \\
thinning = 3 & R.Time & - & - & - & - & - & - & - & - & - \\
\midrule
\multirow{2}{*}{\texttt{meanvarTMD}} & Median & 0.118 & 4.102 & 49.189 & 3.781 & 48.681 & 367.243 & 21.209 & 157.179 & 1215.630 \\
& R.Time & 0.375 & 4.861 & 12.170 & 4.467 & 12.037 & 13.015 & 10.746 & 14.056 & 14.366 \\
\midrule
\multirow{2}{*}{\texttt{TT.moment}} & Median & 7.452 & 24.027 & 94.408 & 62.026 & 202.704 & 789.641 & 242.701 & 800.360 & 3081.367 \\
& R.Time & 23.767 & 28.0473 & 23.358 & 73.279 & 50.122 & 27.984 & 122.974 & 71.574 & 36.414 \\
\bottomrule
    \end{tabular*}
\end{table}


\section{Application on Spatial Model for Censored Data}\label{application}


In this application we will consider the Gaussian spatial censored linear (SCL) model defined by \cite{lachos2017influence} and \cite{ordonez2018geostatistical}. In this model, the data is generated from $\Z = \X\bbeta + \bxi$, with $\bxi \sim \mathcal{N}_p(\textbf{0}, \bSigma)$ and $\bSigma = [\mathrm{Cov}(\as_i,\as_j)] = \sigma^2\textbf{R}(\phi) + \tau^2\textbf{I}_p$. It also has the particularity that the response variable $\Z$ is not fully observed. Instead, it is observed $V_i$ and $C_i$ at each location, for $i = 1, \ldots, p$, where $C_i=0$ and $V_i = Z_i$ for an uncensored observation $Z_i$, and if $C_i = 1$ and $V_i = [V_{1i}, V_{2i}]$ if  $Z_i$ is censored or missing. Because of the difficulties in working directly with the observed likelihood function, \cite{lachos2017influence} suggested using an EM-type algorithm to obtain the ML estimates of $\btheta$ considering a parameterization $\bSigma = \sigma^2\bPsi$, with $\bPsi = \textbf{R}(\phi) + \nu^2\textbf{I}_n$ and $\nu^2 = \tau^2/\sigma^2$, to help with the identifiability of the parameters. See also \cite{diggle2007springer}. Hence, the EM algorithm works as it follows:
\begin{itemize}[leftmargin=0.4cm]
\item \textbf{E-step}: Let $\widehat{\btheta}^{(k)}$ be the current estimate of $\btheta$, then the conditional expectation of the complete-data log-likelihood without the constant is
\begin{equation*}
Q_{k}(\btheta) = \E\left(\ell(\btheta|\Z_c) | \textbf{V}, \textbf{C}, \widehat{\btheta}^{(k)}\right) = -\frac{1}{2}\left[ \log|\bPsi| + n\log\sigma^2 + \frac{1}{\sigma^2} \widehat{A}^{(k)} \right],
\end{equation*}
where $\widehat{A}^{(k)} = \mathrm{tr}(\widehat{\Z\Z^\top}^{(k)} \bPsi^{-1}) - 2 \widehat{\Z}^{(k)\top}\bPsi^{-1}\X\bbeta + \bbeta^\top\X^\top\bPsi^{-1}\X\bbeta$. Therefore, the E-step reduces only to the computation of $\widehat{\Z\Z^\top}^{(k)} = \E\left(\Z\Z^\top |\textbf{V},\textbf{C},\widehat{\btheta}^{(k)}\right)$ and $\widehat{\Z}^{(k)} = \E\left(\Z | \textbf{V}, \textbf{C}, \widehat{\btheta}^{(k)}\right)$. In the traditional EM algorithm, we should now evaluate the conditional expectations, which is possible through the  \textsf{R} packages \textbf{tmvtnorm} or \textbf{MomTrunc}, but it is computationally expensive when the proportion of censored observations is non-negligible. An alternative is to use the MCEM algorithm, which approximates the conditional expectations by using MC integration. For the SCL model, the MCE-step is performed by estimating $\widehat{\Z\Z}^\top$ and $\widehat{\Z}$ through Algorithm \ref{alg2}.

\item \textbf{M-step}: The conditional maximization step is carried out, and $\widehat{\btheta}^{(k)}$ is updated by maximizing $\widehat{Q}_k(\btheta)$ over $\btheta$ to obtain a new estimate $\widehat{\btheta}^{(k+1)}$, which leads to the expressions:
\begin{eqnarray*}
\widehat{\bbeta}^{(k+1)} &=& \left(\X^\top \widehat{\bPsi}^{-1(k)} \X \right)^{-1} \X^\top \widehat{\bPsi}^{-1(k)} \widehat{\Z}^{(k)}, \\
\widehat{\sigma}^{2(k+1)} &=& \frac{1}{n}\left[ \mathrm{tr}\left(\widehat{\Z\Z^\top}^{(k)}\widehat{\bPsi}^{-1(k)}\right) - 2\widehat{\Z^\top}^{(k)}\widehat{\bPsi}^{-1(k)}\X\widehat{\bbeta}^{(k+1)} + \widehat{\bbeta}^{\top(k+1)} \X^\top \widehat{\bPsi}^{-1(k)} \X\widehat{\bbeta}^{(k+1)} \right],\\
\widehat{\balpha}^{(k+1)} &=& \amax{\balpha\in\mathbb{R}^+\times\mathbb{R}^+} \left( -\frac{1}{2}\log|\bPsi| -\frac{1}{2\widehat{\sigma}^{2(k+1)}} \left[ \mathrm{tr}\left(\widehat{\Z\Z^\top}^{(k)} \bPsi^{-1}\right) - 2\widehat{\Z^\top}^{(k)} \bPsi^{-1}\X\widehat{\bbeta}^{(k+1)} \right.\right. \\
 & & \left. \left. + \widehat{\bbeta}^{\top(k+1)} \X^\top \bPsi^{-1} \X\widehat{\bbeta}^{(k+1)} \right] \right), 
\end{eqnarray*}
with $\balpha=(\phi,\nu^2)^\top$. Note that $\widehat{\tau}^2$ can be recovered by $\widehat{\tau}^{2(k+1)} = \widehat{\nu}^{2(k+1)} \widehat{\sigma}^{2(k+1)}$. An efficient M-step can
be easily accomplished by using, for instance, the \textbf{roptim} package \citep{panroptim}. In general, the estimates of $\btheta$ may vary slightly around the maximum, with a variability depending on the sample size used in the approximation. Hence, one possible final estimate of the parameters may be computed as the mean of the estimates after applying a burn-in and a thinning process.
\end{itemize}

\subsection{Missouri Dioxin Contamination Data}

The proposed MCEM algorithm will be applied to analyze the Missouri dioxin contamination dataset available in \texttt{CensSpatial} package. The dataset contains 127 observations distributed in an area of $3600 \times 65$ $m^2$ on the shoulders of a country road located in Missouri, U.S.A. The observations correspond to the level of contamination by dioxin (2,3,7,8-tetrachlorodibenzo-p-dioxin or TCDD) at sampled points along the road, where 43\% of the observations (55 sites) were censored, falling below some limit of detection, which ranges from 0.10 to 0.79 mg/kg. The spatial directions are the $x$-direction (measured in 1/100 $ft$) and the $y$-direction (in $ft$). Please, refer to \cite{fridley2007data} for more details.

This dataset was firstly analyzed by \cite{zirschky1986geostatistical}, who concluded that data appeared to be log-normally distributed. Hence, we fit the model $\log(Z_i) = \beta_0 + \xi_i$, $i=1,\ldots, 127$. The model parameters were estimated using the MCEM algorithm and compared with the estimates from the SAEM and EM algorithm. All methods were performed using 500 iterations and an exponential correlation function to take into account the variation between spatial points. For the MCEM algorithm, we evaluated four cases; in one of those scenarios, it was considered linearly increasing sample sizes between 100 and 1000. Other scenarios considered constant sample sizes of 20, 5000, and $10^5$. In order to use the SAEM algorithm, we set two configurations; one draws points using the \texttt{rmvtnorm} function (from package \textbf{tmvtnorm}), and the optimization procedure via \texttt{optimx} function \citep{nash2020package}. This method is available in the \textbf{CensSpatial} package, and from now on, we refer to this algorithm by SAEM. The second one draws points using the proposed slice sampler, while the \textsf{R} function \texttt{roptim} executes the optimization procedure. We will refer to the latter as SAEM-SS. Lastly, moments were computed using the \textbf{MomTrunc} package for the EM algorithm.  The functions used to estimate the parameters via MCEM, SAEM-SS, and EM are available in the \textbf{RcppCensSpatial} package.

\begin{table}[ht]
\centering
\small
\caption{Missouri data - ML estimates and information criteria (AIC and BIC) obtained through MCEM, SAEM and EM algorithms considering the exponential correlation function.} \label{app_est}
\begin{tabular*}{\textwidth}{l@{\extracolsep{\fill}}cccccccccc}
\toprule
Algorithm & $n$ & $c$ & $\beta_0$ & $\sigma^2$ & $\phi$ & $\tau^2$ & Log-likelihood & AIC & BIC & Time (min) \\
\bottomrule
\multirow{4}{*}{MCEM} & 20 & & -2.355 & 6.577 & 14.702 & 0.213 & -143.128 & 294.257 & 305.633 & 0.936 \\
& $10^2$ - $10^3$ & - & -2.402 & 6.808 & 15.095 & 0.207 & -143.108 & 294.216 & 305.592 & 1.819 \\
& $5000$ & - & -2.410 & 6.847 & 15.076 & 0.206 & -143.095 & 294.191 & 305.568 & 10.206 \\
& $10^5$ & - & -2.408 & 6.845 & 15.053 & 0.205 & -143.136 & 294.272 & 305.649 & 185.341 \\
\midrule
SAEM-SS & 20 & 0.25 & -2.332 & 6.312 & 15.109 & 0.214 & -143.153 & 294.307 & 305.683 & 0.676 \\ 
\midrule
\multirow{2}{*}{SAEM} & 20 & 0.25 & -2.014 & 4.858 & 14.206 & 0.245 & -143.840 & 295.681 & 307.057 & 6.079 \\
& $10^5$ & 1.00 & -2.010 & 4.829 & 14.136 & 0.245 & -143.865 & 295.729 & 307.106 & 9151.149 \\
\midrule
EM & - & - & -2.417 & 6.888 & 15.092 & 0.206 & -143.122 & 294.244 & 305.620 & 1661.472 \\
\bottomrule
\end{tabular*}
\end{table}

The results of the ML estimates are shown in Table \ref{app_est}, where $n$ is the number of samples considered to approximate the conditional mean, and $c$ indicates the percentage of iterations without memory in the SAEM algorithm \citep{lachos2017influence, ordonez2018geostatistical}. Final estimates for the MCEM and EM methods were computed as the mean of the estimates at each iteration after applying a burn-in of 250 and thinning of 3 observations, while the SAEM and SAEM-SS estimates we only considered the estimates at the last iteration. We see that estimates obtained from SAEM-SS are similar to MCEM estimates for $n=20$, while the estimates for the EM algorithm are similar to MCEM estimates with $n=5000$. The estimates obtained through MCEM and EM for the regression coefficient $\beta_0$ were -2.400, while the SAEM algorithm estimated this parameter equal to -2.010. Regarding to the spatial scaling parameter $\phi$, it was around 15.05 and 14.10 for the MCEM and SAEM algorithms, respectively. These values imply that for distances greater than 45 and 42 feet, respectively, the correlation between two observations falls to less than 0.05. The estimates achieved from MCEM and SAEM methods for the partial sill $\sigma^2$ and the nugget effect $\tau^2$ suggest that 97\% and 95\% of the variability in data is explained by the spatial process, respectively. This table also shows the maximized log-likelihood value, information criteria AIC and BIC, and the running time in minutes. Based on information criteria, we can conclude that MCEM with $n=5000$ best fits the Missouri dioxin contamination data. Furthermore, it does not seem necessary to consider sample sizes as large as $n = 10^5$ because that configuration does not gain the precision of the estimates and is more time-consuming.

\begin{figure}[ht]
\centering
\caption{Missouri data - Convergence of the parameter estimates via EM, MCEM, and SAEM algorithm.}
\label{app_convergence}
\textit{a. EM algorithm}
\includegraphics[scale=.68]{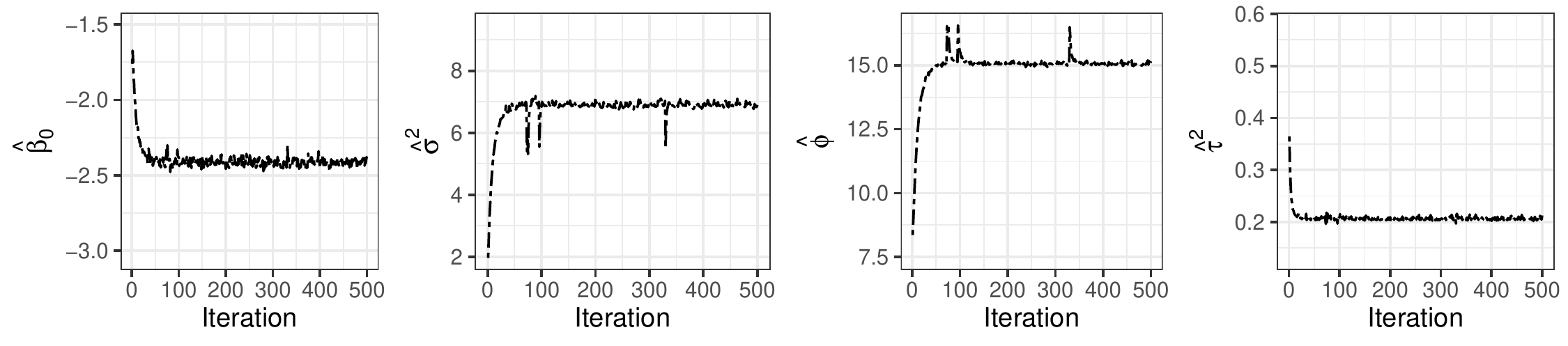}

\textit{b. MCEM algorithm}
\includegraphics[scale=.68]{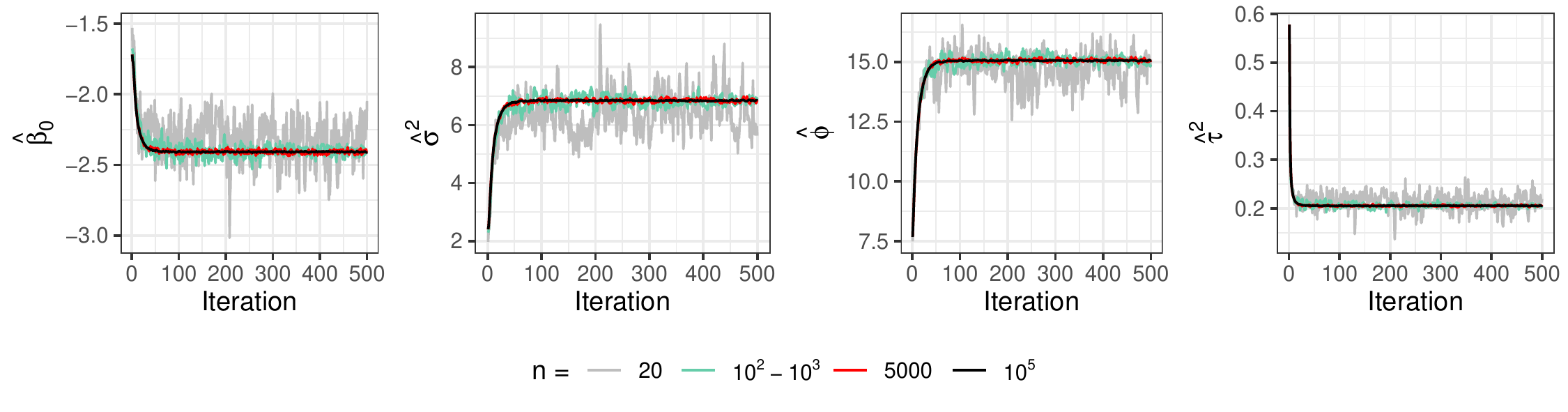}

\textit{c. SAEM algorithm}
\includegraphics[scale=.68]{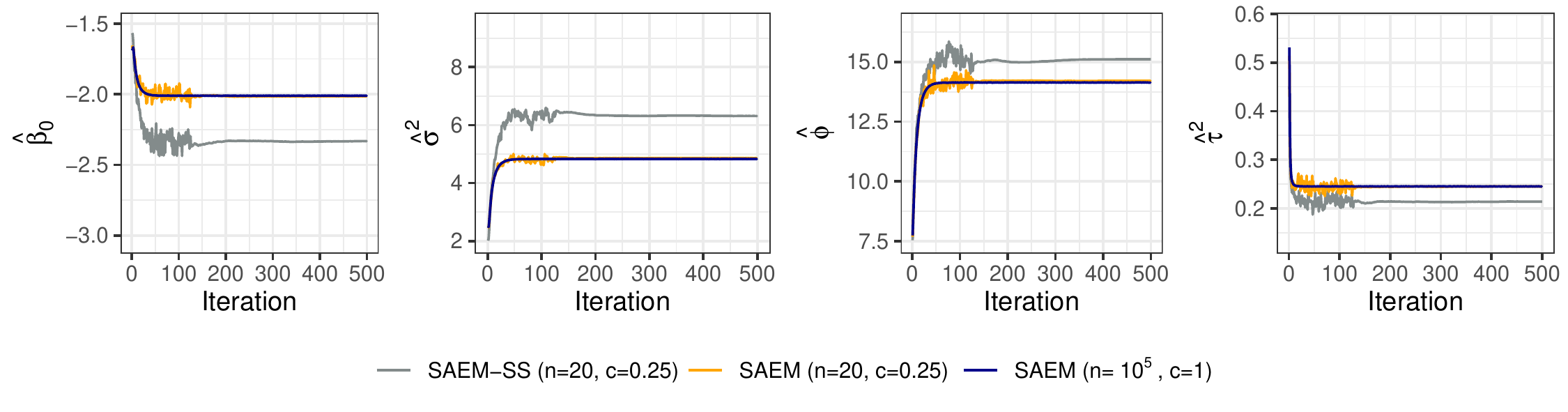}
\end{figure}

Figure \ref{app_convergence} shows the convergence graphs of the parameter estimates achieved from MCEM, SAEM-SS, SAEM, and EM algorithms. Notice that the variability in the estimates for MCEM decreases when the sample size increases from 100 to 1000 (aquamarine line). As expected, the estimates obtained from MCEM with $n=20$ (gray line) present more variability than the other three scenarios in which we considered larger sample sizes, while MCEM with $n=10^5$ (black line) reported the lowest variability in the estimates. The estimates of the parameters computed through the EM algorithm present more variability than the estimates from MCEM with $n=5000$ (red line), probably due to computational stability of the numerical methods involved in the \textbf{MomTrunc} package; this is why we decided to consider a burn-in and thinning procedure to compute the EM final estimates.


\section{Conclusions}\label{conclusions}


This work describes an algorithm to generate random numbers from members of the truncated elliptical family of distributions with a strictly decreasing density generating function through a slice sampling algorithm and a Gibbs sampler step. In addition, we presented an efficient approach to approximate the first and the second moment for these distributions. We briefly introduce the functions available in our \textsf{R} package \textbf{relliptical} in order to perform sample generating and estimation of the first two moments. Simulation studies were performed to investigate the properties of estimates and the robustness of our algorithm. Moreover, we compared our approach with others available in the \textsf{R} software (only for the normal and Student-$t$ case), where we showed that our approach over-performed others in terms of precision and computational time. We illustrate the usefulness of truncated moments on the Missouri dioxin contamination dataset, where a spatial model for censored data was fitted. 

Future extensions of the work include the extension of this method to the context of asymmetric multivariate elliptical distributions, so the fast computation of their truncated moments may lead the way to proposed more flexible and robust models relating censored models for mixed-effects models, longitudinal data, spatial models, among others. Finally, results presented in this paper can be reproduced through the \textsf{R} package \textbf{relliptical}, which is available at CRAN for download. 


\section*{Acknowledgements}

The research of Katherine A. L. Valeriano was supported by CAPES. Larissa A. Matos acknowledges support from FAPESP-Brazil (Grant 2020/16713-0). 

\bibliographystyle{chicago}  
\bibliography{references}  

\newpage
\appendix
\section{Extra simulation results} 

This appendix contains additional information about the simulation results.

\subsection{Sample autocorrelation} \label{AppA}

\begin{figure}[ht]
    \centering
    \caption{Sample autocorrelation plots of $X_1$ and $X_2$ sampled from the bivariate truncated elliptical distributions in Figure \ref{fig_rng}.} \label{fig_acf}
    \vspace*{0.1cm}
    \includegraphics[scale=.68]{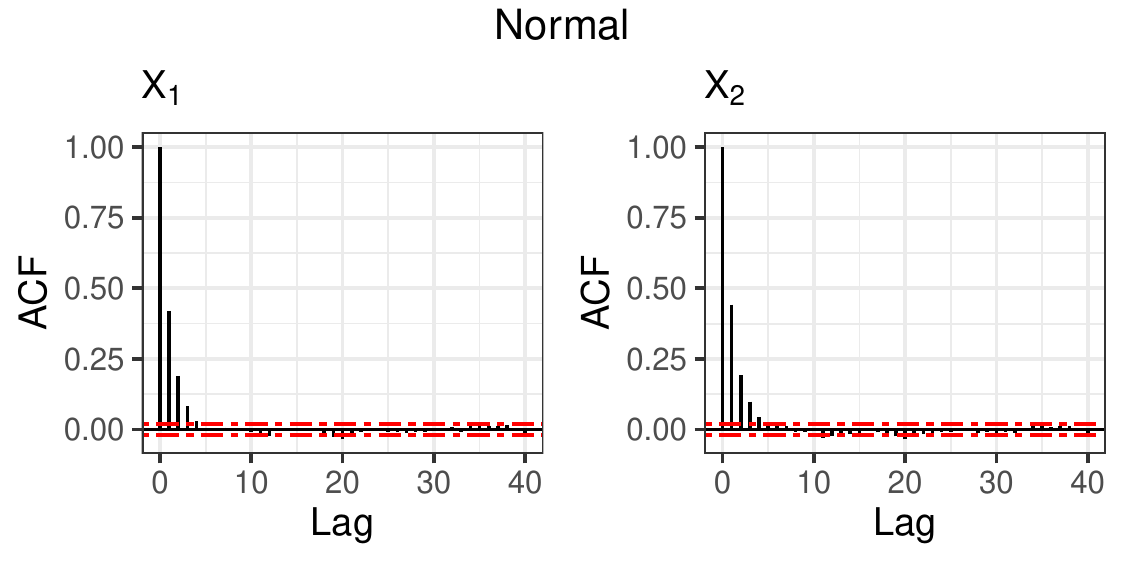}
    \includegraphics[scale=.68]{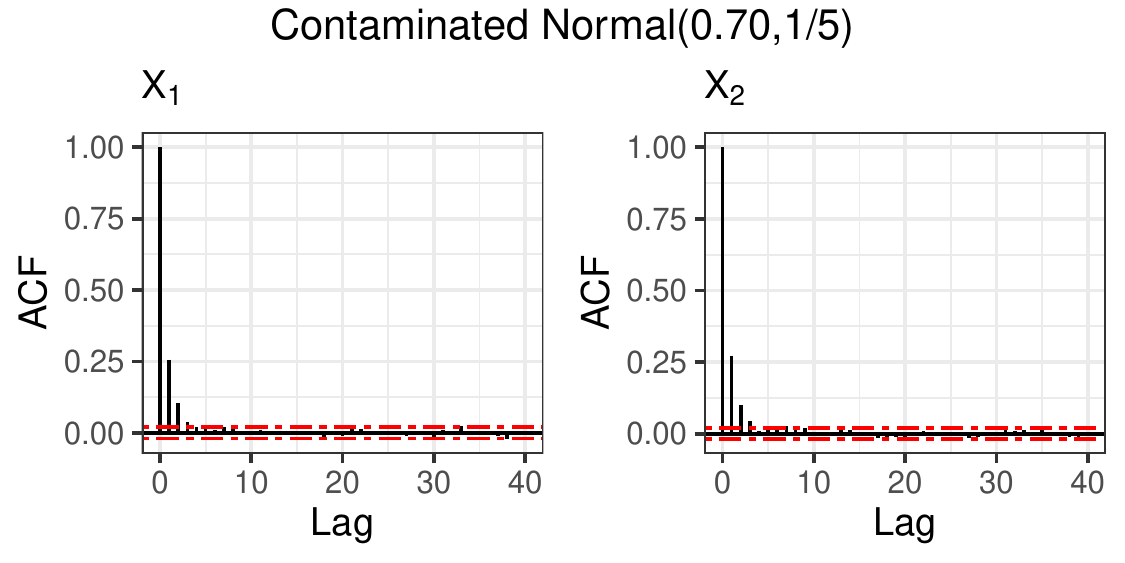}
    \includegraphics[scale=.68]{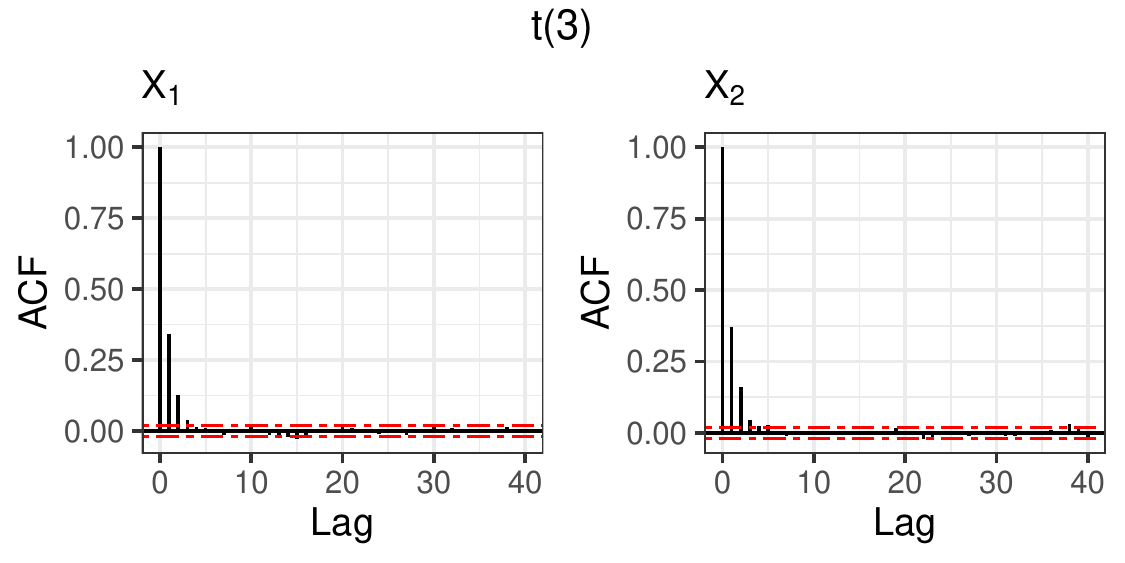}
    \includegraphics[scale=.68]{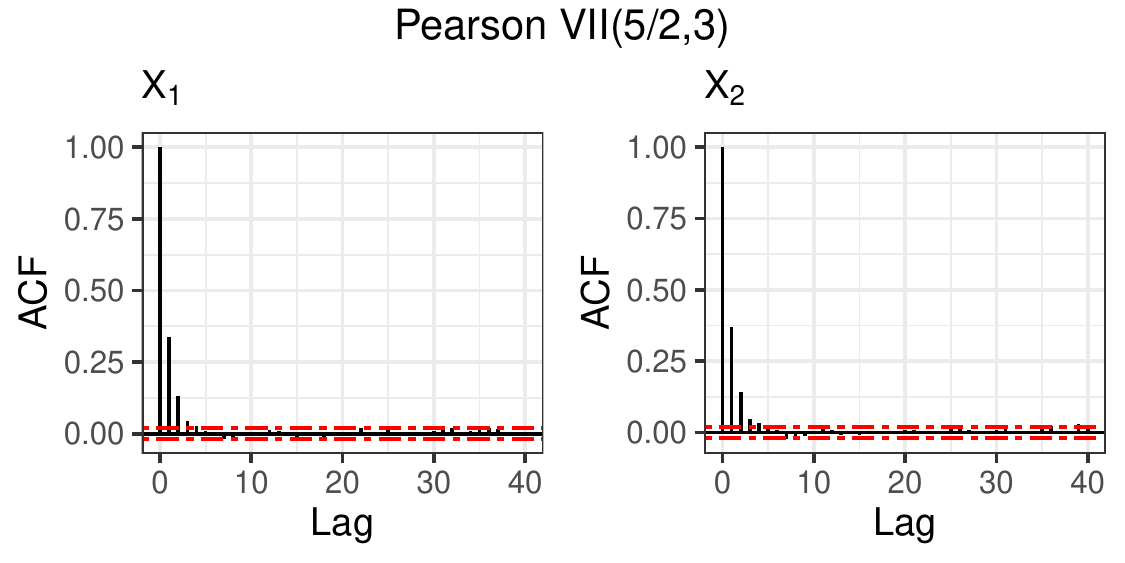}
    \includegraphics[scale=.68]{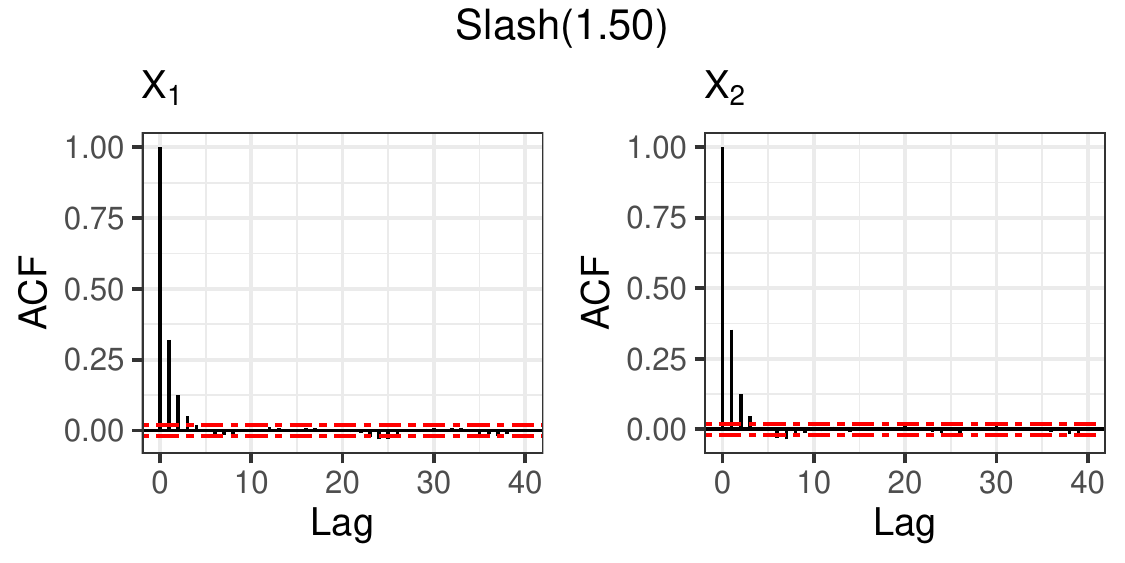}
    \includegraphics[scale=.68]{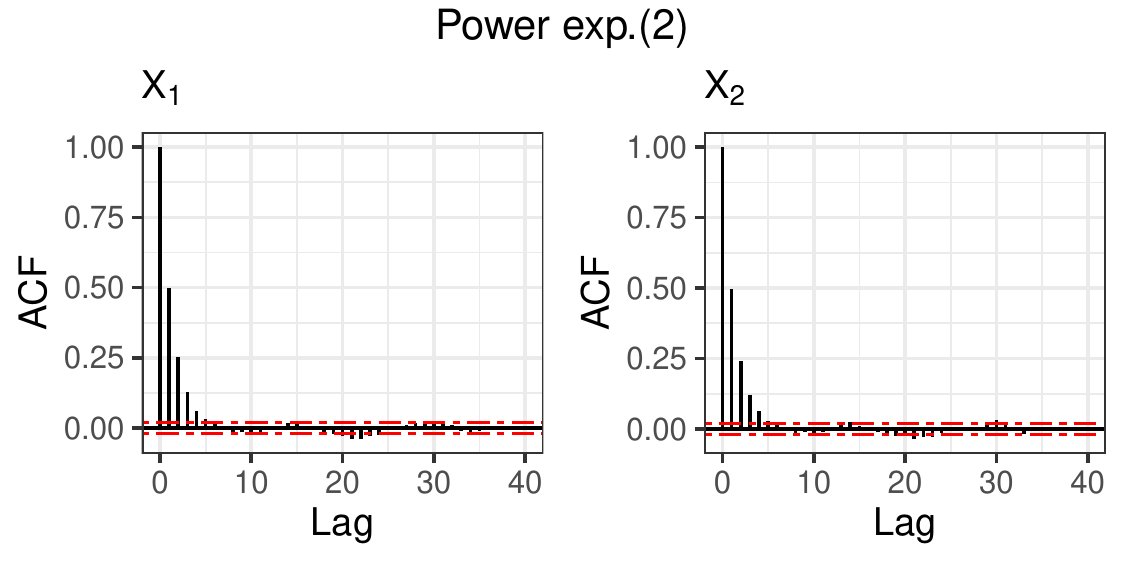}
    \includegraphics[scale=.68]{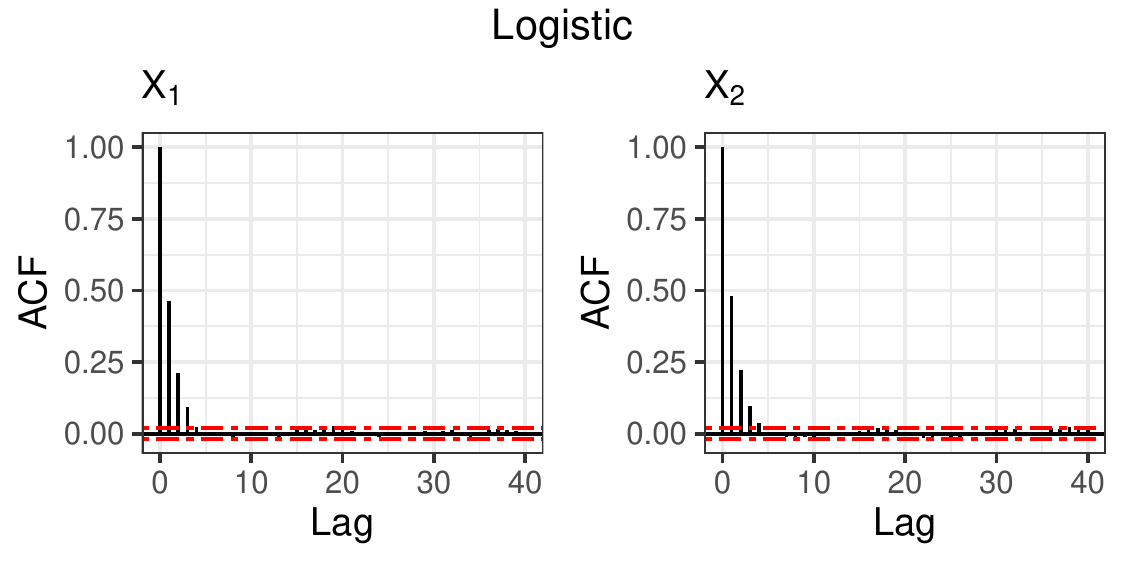}
    \includegraphics[scale=.68]{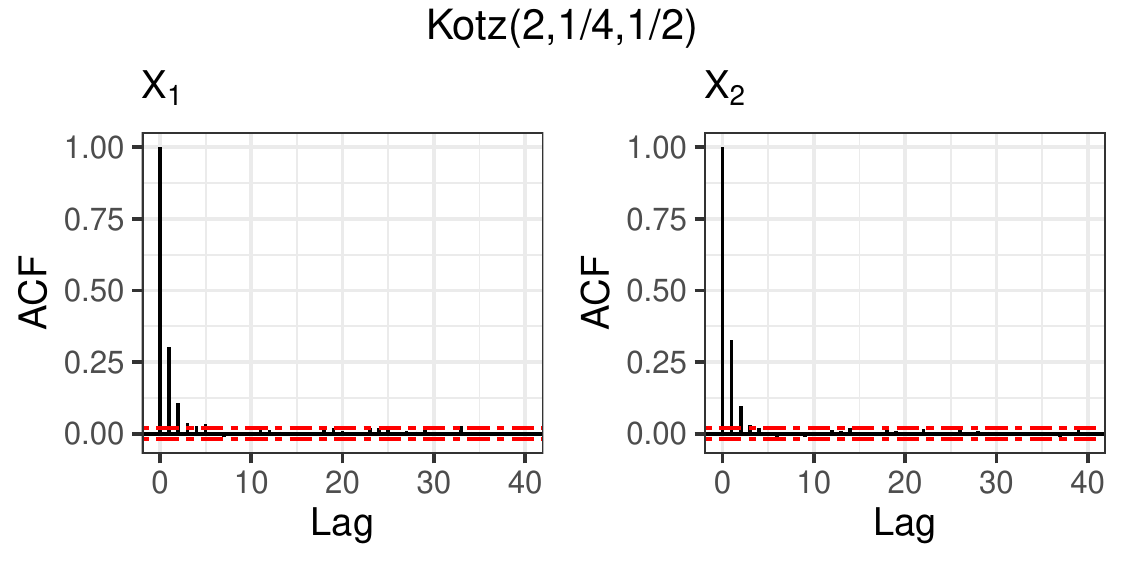}
\end{figure}

\subsection{Running Time to Compute Moments from Truncated Distributions} \label{AppAx}

In this section, a complementary study of Example 2 (Section \ref{meanvar}) was conducted to examine the execution time required for our method in order to estimate the first two moments and the variance-covariance matrix of a $p$-variate random vector considering different distributions of the truncated elliptical family, with $p=50$ and $100$. As in Example 2, for each case we consider a 10\%, 20\%, and 40\% of doubly truncated variables.

Table \ref{time} shows the median of the execution time (in seconds) needed for function \texttt{mvtelliptical} to compute the first two moments and the covariance matrix. We considered a TMVN, a truncated contaminated normal with $\nu=1/2$ and $\rho=1/5$, a truncated Pearson VII with parameters $m=55$ and $\nu=3$, a truncated slash with $\nu=2$ degrees of freedom, and a truncated power exponential distribution with kurtosis $\beta=1/2$. For each case, our method was applied setting $n=10^4$ and $10^5$ with a thinning$=3$. Notice that the time needed by the algorithm for TMVN, TMVT, and truncated Pearson VII distributions are similar and depend only on the number truncated variables and samples used in the approximation. Our method requires more time to compute moments from the truncated contaminated normal distribution when compared to the latter results. This is because the algorithm uses a numerical method to calculate the inverse of the dgf. Besides, it is interesting noting that there is no time difference between computing the moments for a truncated slash distribution with five or ten doubly truncated variables. This occurs since the function used to approximate the integral on the dgf is more time-consuming when $\nu + p/2 - 1$ is not an integer. Finally, the computation of the moments for the truncated power exponential distribution required approximately the same time for random vectors of equal length regardless of the number of doubly truncated variables. For this case, the method samples values for the whole vector, leading to no time difference.

\begin{table}[ht]
\centering
\caption{Median of the execution time (in seconds) based on 100 simulations.} \label{time}
\begin{tabular*}{\textwidth}{l@{\hspace{\tabcolsep}\extracolsep{\fill}}ccccccc}
\toprule
\multirow{2}{*}{Distribution ($\nu$)} & \multirow{2}{*}{Sample size} & \multicolumn{3}{c}{$p=50$} & \multicolumn{3}{c}{$p=100$} \\
\cmidrule{3-5}  \cmidrule{6-8}
& & 10\% & 20\% & 40\% & 10\% & 20\% & 40\% \\
\bottomrule
\multirow{2}{*}{Normal} & $10^4$ & 0.028 & 0.083 & 0.399 & 0.081 & 0.400 & 2.888 \\
& $10^5$ & 0.285 & 0.840 & 3.999 & 0.805 & 4.003 & 28.892 \\
\midrule
Contaminated & $10^4$ & 0.071 & 0.118 & 0.440 & 0.120 & 0.442 & 2.928 \\
Normal (1/2, 1/5) & $10^5$ & 0.706 & 1.180 & 4.405 & 1.192 & 4.415 & 29.286 \\
\midrule
\multirow{2}{*}{Pearson VII (55, 3)} & $10^4$ & 0.031 & 0.083 & 0.403 & 0.084 & 0.403 & 2.891 \\
& $10^5$ & 0.309 & 0.838 & 4.030 & 0.839 & 4.036 & 28.944 \\
\midrule
\multirow{2}{*}{Slash (2)} & $10^4$ & 0.202 & 0.202 & 0.548 & 0.200 & 0.549 & 3.113 \\
& $10^5$ & 2.020 & 2.026 & 5.481 & 1.997 & 5.489 & 31.160 \\
\midrule
Power & $10^4$ & 5.101 & 5.095 & 5.096 & 41.870 & 41.858 & 41.864 \\
Exponential (1/2) & $10^5$ & 51.038 & 51.013 & 50.999 & 418.675 & 418.604 & 418.651 \\
\bottomrule
    \end{tabular*}
\end{table}


\section{The Multivariate Pearson VII Distribution} \label{pvii}

\subsection{Marginal and conditional distributions}

A random variable $\X\in\mathbb{R}^p$ is said to have a multivariate Pearson VII distribution with location parameter $\bmu\in\mathbb{R}^p$, positive-definite scale matrix $\bSigma\in\mathbb{R}^{p\times p}$, extra parameters $m>p/2$ and $\nu>0$, if its pdf is given by
\begin{eqnarray*}
f_{\X}(\x) = \frac{\Gamma(m)}{(\pi \nu)^{p/2} \Gamma(m - p/2)} |\bSigma|^{-1/2} \left( 1 + \frac{1}{\nu}(\x-\bmu)^\top \bSigma^{-1} (\x-\bmu)\right)^{-m}, \quad \x\in\mathbb{R}^p.
\end{eqnarray*}

The random vector $\X$ can also be represented as a scale mixture of normal (SMN) distributions, i.e., $\X = \bmu + U^{-1/2}\Z$, where $\Z$ has a $p$-variate normal distribution with mean $\textbf{0}\in\mathbb{R}^p$ and variance-covariance matrix $\bSigma\in\mathbb{R}^{p\times p}$. Here, $U$ follows Gamma distribution with scale parameter $m - p/2$ and rate parameter $\nu/2$, where $\Z$ is independent of $U$. This implies that
\begin{equation*}
\X|(U = u) \sim \mathcal{N}_p(\bmu, u^{-1}\bSigma) \quad \mbox{and} \quad U\sim \mathcal{G}(m-p/2, \nu/2).    
\end{equation*}
Therefore, the mean and the variance-covariance matrix of $\X$ are
\begin{eqnarray*}
\mathbb{E}(\X) &=& \mathbb{E}(\mathbb{E}(\X|U)) = \mathbb{E}(\bmu) = \bmu, \quad m > \frac{p + 1}{2}.\\
\mathrm{\mathrm{C}ov}(\X) &=& \mathrm{Cov}(\mathbb{E}(\X|U)) + \mathbb{E}(\mathrm{Cov}(\X|U)) = \mathbb{E}(U^{-1})\bSigma = \frac{\nu}{2m - p - 2} \bSigma, \quad m > \frac{p + 2}{2}. \hspace{2cm}
\end{eqnarray*}

Now suppose that the vector $\X$ is partitioned into two random vectors $\X_1\in \mathbb{R}^{p_1}$ and $\X_2\in\mathbb{R}^{p_2}$, such that $p=p_1+p_2$, and consider the partition of $\bmu$ and $\bSigma$ used in Proposition \ref{remark1}, i.e., 
\begin{eqnarray*}
\X = \left( \begin{array}{c} \X_1 \\ \X_2\end{array}\right), \quad \bmu = \left( \begin{array}{c} \bmu_1 \\ \bmu_2\end{array}\right) \quad \mbox{and} \quad \bSigma = \left( \begin{array}{cc} \bSigma_{11} & \bSigma_{12} \\ \bSigma_{21} & \bSigma_{22} \end{array}\right).
\end{eqnarray*} 

\noindent First, notice that $(\X-\bmu)^\top\bSigma^{-1}(\X-\bmu) = \delta_1(\X_1) + \delta_{2.1}(\X_{2.1})$, where $\delta_1(\X_1) =  (\X_1-\bmu_1)^\top\bSigma_{11}^{-1}(\X_1-\bmu_1)$, $\delta_{2.1}(\X_{2.1}) = (\X_2 - \bmu_{2.1})^\top\bSigma_{2.1}^{-1}(\X_2 - \bmu_{2.1})$, 
$\bmu_{2.1} = \bmu_2 + \bSigma_{21}\bSigma_{11}^{-1}(\X_1 - \bmu_1)$ and $\bSigma_{2.1}=\bSigma_{22} - \bSigma_{21}\bSigma_{11}^{-1}\bSigma_{12}$. By the results above, the marginal pdf of $\X_1$ is given by
\begin{eqnarray*}
f_{\X_1}(\x_1) &=& \int_{\mathbb{R}^{p_2}} f_{\X}(\X) \dr\x_2 = \frac{\Gamma(m)}{(\pi\nu)^{p/2}\Gamma(m-p/2)} |\bSigma|^{-1/2} \int_{\mathbb{R}^{p_2}} \left( 1 + \frac{\delta_1(\x_1)}{\nu} + \frac{\delta_{2.1}(\x_{2.1})}{\nu} \right)^{-m} \dr\x_2 \\
&=& \frac{\Gamma(m)}{(\pi\nu)^{p/2}\Gamma(m-p/2)} |\bSigma|^{-1/2} \left( 1 + \frac{\delta_1(\x_1)}{\nu}\right)^{-m} \int_{\mathbb{R}^{p_2}} \left( 1 + \frac{ \delta_{2.1}(\x_{2.1}) }{\nu + \delta_1(\x_1)}\right)^{-m} \dr\x_2 \\
&=& \frac{\Gamma(m -  p_2/2)}{(\pi\nu)^{p_1/2} \Gamma(m - p/2)} |\bSigma_{11}|^{-1/2} \left( 1 + \frac{\delta_1(\x_1)}{\nu}\right)^{-(m-p_2/2)}, \quad \x_1\in\mathbb{R}^{p_1}.
\end{eqnarray*}
Hence, the marginal distribution of $\X_1$ is also Pearson VII distributed with parameters $\bmu_1$, $\bSigma_{11}$, $m-p_2/2$ and $\nu$, i.e., $\X_1\sim \mbox{PVII}_{p_1}(\bmu_1, \bSigma_{11}, m-p_2/2, \nu)$. On the other hand, the conditional pdf of $\X_2|(\X_1 = \x_1)$ is given by
\begin{eqnarray*}
f_{\X_2|\X_1}(\x_2|\x_1) &=& \frac{f_{\X}(\x_1, x_2)}{f_{\X_1}(\x_1)} \\
&=& \frac{\Gamma(m) |\bSigma_{2.1}|^{-1/2}}{(\pi (\nu+\delta_1(\x_1)))^{p_2/2} \Gamma(m-p_2/2)} \left( 1 + \frac{\delta_{2.1}(\x_{2.1})}{\nu + \delta_1(\x_1)}\right)^{-m}, \quad  \x_1\in\mathbb{R}^{p_1}, \, \x_2\in\mathbb{R}^{p_2}.
\end{eqnarray*}
Therefore, the conditional distribution has also a Pearson VII distribution with parameters $\bmu_{2.1}$, $\bSigma_{2.1}$, $m$ and $\nu + \delta_1(\x_1)$, i.e., $\X_2|(\X_1=\x_1) \sim \mbox{PVII}_{p_2}(\bmu_{2.1}, \bSigma_{2.1}, m, \nu+\delta_1(\x_1))$.

\subsection{Existence of its truncated moments}

Let $\X \sim \mbox{PVII}_p(\bmu, \bSigma, m, \nu), m > p/2, \nu > 0$, and let ${A} \subseteq \mathbb{R}^p$ be a truncation region of interest. Then, the expectation and the variance-covariance matrix of $\X$ given $\X \in {A}$ exist in the following cases:

\begin{itemize}[leftmargin=0.40cm]
\item If ${A} = \mathbb{R}^p$ or ${A}$ is unbounded, the vector is not truncated at all, so the expectation exists for $m > (p+1)/2$ and the covariance matrix exists for $m > (p+2)/2$, as usual.

\item If ${A}$ is bounded (all truncation points are finite), then $\mathbb{E}(\X | \X\in {A}) < \infty$ and $\mathrm{Cov}(\X| \X\in {A}) < \infty$ for all $m > p/2$, since the distribution is bounded. 

\item If $\X$ can be partitioned into two random variables $\X_1\in\mathbb{R}^{p_1}$ and $\X_2\in\mathbb{R}^{p_2}$ such that the truncation region associated to $\X_1$ (say, ${A}_1$) is bounded, from the last item we have $\mathbb{E}(\X_1| \X \in {A})$ and $\mathrm{Cov}(\X_1 | \X \in {A})$ exist for all $m > p/2$ and $\nu > 0$. On the other hand, it follow from Fubini's theorem that $\mathbb{E}(\X_2 | \X \in {A})$ will exist if and only if $\mathbb{E}(\X_2| \X_1)$ exists; this occurs for all $m > (p_2 + 1)/2$. Note that $\mathbb{E}(\X_2|\X_1) < \infty$ also implies that $\mathrm{Cov}(\X_1,\X_2| \X\in {A}) < \infty$. Additionally, $\mathrm{Cov}(\X_2 | \X\in {A})$ exists if and only if $\mathrm{Cov}(\X_2| \X_1) < \infty$, which holds for all $m > (p_2 + 2)/2$. 
\end{itemize}

\begin{remark}
It is equivalent to say that $\mathbb{E}(\X | \X\in {A})$ exists for all $m$, if at least one
dimension containing a finite limit exists. Besides, if at least two dimensions containing finite limits exist, we have that $\mathrm{Cov}(\X | \X\in {A})$ exists for all $m > p/2$.

\end{remark}

\begin{figure}[ht]
\centering
\caption{Trace plots of the evolution of the MC estimates for the mean and variance-covariance elements of $\X \mid (\X \in A)$ under scenarios a), b) and c). The red dashed line represents the true estimated value computed using numerical methods} \label{tpvii}
a. Two non-truncated variables, parameters $m=2$ and $\nu=1$. 
\includegraphics[scale=.73]{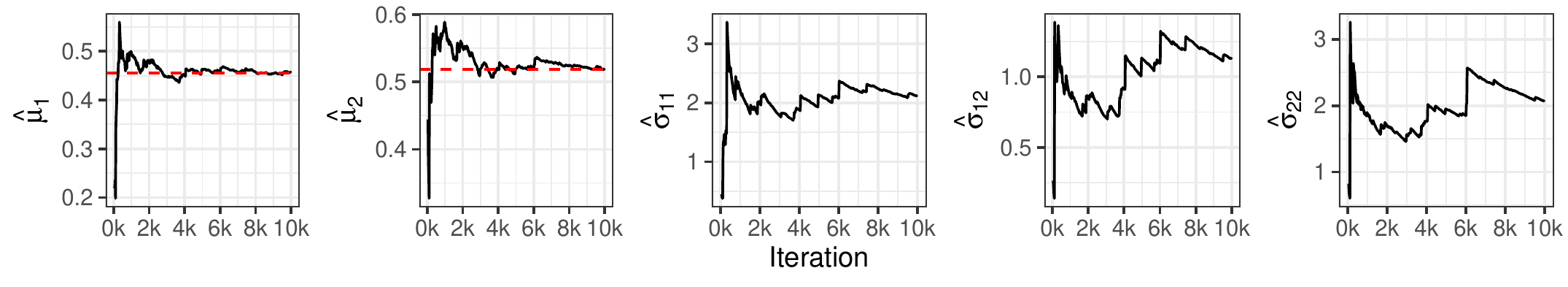}

b. One doubly truncated variable, parameters $m=1.40$ and $\nu=1$.
\includegraphics[scale=.73]{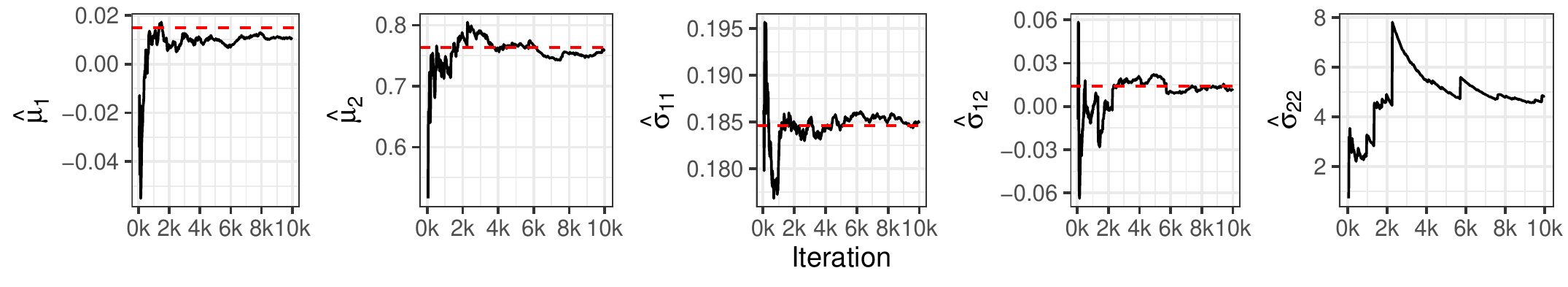}

c. One doubly truncated variable, parameters $m=2$ and $\nu=1$.
\includegraphics[scale=.73]{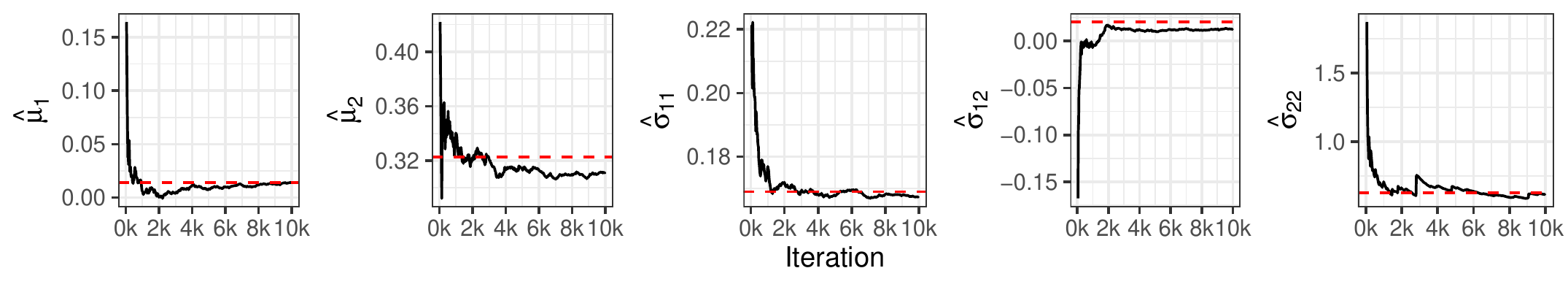}
\end{figure}

In order to illustrate the result, consider $\X \sim \mbox{PVII}_2(\bmu, \bSigma, m, \nu)$, with $\nu = 1$, $\bmu = \textbf{0}$, and $\bSigma = \left( \begin{array}{cc} 1 & 0.20 \\ 0.20 & 1\end{array}\right)$. We are interesting to observe what happens with the elements of $\mathbb{E}(\X | \X\in {A})$ and $\mathrm{Cov}(\X | \X \in {A})$ for ${A} = \{\x\in \mathbb{R}^2 : \textbf{a}< \x < \textbf{b}\}$ in the following three scenarios: 
\begin{itemize}[leftmargin=0.40cm]
\item[a)] $m = 2$, $\textbf{a} = (-0.80, -0.60)^\top$, $\textbf{b}= (\infty, \infty)^\top$; 
\item[b)] $m = 1.40$, $\textbf{a} = (-0.80, -0.60)^\top$, $\textbf{b} = (0.80, \infty)^\top$; 
\item[c)] $m = 2$, $\textbf{a} = (-0.80, -0.60)^\top$, $\textbf{b} = (0.80, \infty)^\top$. 
\end{itemize}
Figure \ref{tpvii} displays the trace evolution of the MC estimates for the mean and variance-covariance elements $\mu_1$, $\mu_2$, $\sigma_{11}$, $\sigma_{12}$ and $\sigma_{22}$ for each case. The red dashed line represents the value for the parameter estimated via MC with $10^6$ samples, and we refer to this value as the ``real value".

For the first case, we have that $(p+1)/2 =3/2 < 2 = m$, then only the first moment exists, i.e., $\mathbb{E}(\X| \X\in A) < \infty$. Therefore, we observe in the first row of Figure \ref{tpvii} that only the estimates of $\mu_1$ and $\mu_2$ converge to their real values as the sample size increase. In the second scenario (middle row), we have that all elements converge except $\sigma_{22}$. This happens because the truncation limits for the first variable are finite and $m > (p_2 + 1)/2 = 1$. In the last case, scenario c), convergence is attained for all parameters, since the condition $m > (p_2 + 2)/2 = 3/2$ holds. Note that even with 2000 MC simulations there exists a significant variability in the chains.

\section{The Multivariate Slash Distribution}\label{slash}

A random vector $\X\in\mathbb{R}^p$ has multivariate slash distribution with location parameter $\bmu\in\mathbb{R}^p$, positive-definite scale matrix $\bSigma\in\mathbb{R}^{p\times p}$ and $\nu>0$ degrees of freedom, denoted by $\X\sim\mbox{SL}_p(\bmu, \bSigma, \nu)$, if its pdf is given by
\begin{equation*}
f_\X(\x) = \nu \int_0^1 u^{\nu-1} \phi_p\left(\x; \bmu, u^{-1}\bSigma \right) \dr u, \quad \x\in\mathbb{R}^p,   
\end{equation*}
where $\phi_p(\x;\bmu,\bSigma)$ is the pdf of a $p$-variate normal distribution with mean $\bmu$ and covariance matrix $\bSigma$. The pdf of a slash distribution can be evaluated through numerical methods, e.g., using the  \textsf{R} function \texttt{integrate}. The random vector $\X$ can also be represented in the family of the SMN distributions, this is, $\X = \bmu + U^{-1/2}\Z$, where the random variables $U$ and $\Z$ are both independent and have $\mathrm{Beta}(\nu,1)$ and $\N_p(\textbf{0},\bSigma)$ distributions, respectively. Therefore, the mean and variance-covariance matrix of the random vector $\X$ are given by
\begin{eqnarray*}
\E(\X) &=& \E\left(\E(\X | U )\right) = \E(\bmu) = \bmu. \\
\mathrm{Cov}(\X) &=& \mathrm{Cov}(\E(\X|U)) + \E(\mathrm{Cov}(\X|U)) = \E(U^{-1})\bSigma = \frac{\nu}{\nu-1}\bSigma, \quad \nu>1. \hspace*{3cm}
\end{eqnarray*}

\noindent Considering a partition in the same manner as used for the Pearson VII distribution, the marginal pdf of $\X_1$ is given by
\begin{eqnarray*}
f_{\X_1}(\x_1) &=& \int_{\mathbb{R}^{p_2}} f_{\X}(\x) d\x_2 = \int_{\mathbb{R}^{p_2}} \nu \int_0^1 u^{\nu - 1} \phi_p\left(\x; \bmu, u^{-1}\bSigma\right) \dr u \, d\x_2 \\
&=& \nu \int_{\mathbb{R}^{p_2}} \int_0^1 u^{\nu-1} \phi_{p_1}\left(\x_1; \bmu_{1}, u^{-1}\bSigma_{11}\right) \phi_{p_2}\left(\x_2; \bmu_{2.1}, u^{-1}\bSigma_{2.1}\right) \dr u \,\dr\x_2 \\
&\stackrel{\mbox{Fubini}}{=}& \nu \int_0^1 u^{\nu-1} \phi_{p_1}\left(\x_1; \bmu_{1}, u^{-1}\bSigma_{11}\right)  \int_{\mathbb{R}^{p_2}} \phi_{p_2}\left(\x_2; \bmu_{2.1}, u^{-1}\bSigma_{2.1}\right) \dr\x_2\,\dr u  \hspace*{3cm}
\end{eqnarray*}
\begin{eqnarray*}
\Rightarrow f_{\X_1}(\x_1) = \nu \int_0^1 u^{\nu-1} \phi_{p_1}\left(\x_1; \bmu_{1}, u^{-1}\bSigma_{11}\right) \dr u. \hspace*{8cm}
\end{eqnarray*}

\noindent Thus, $\X_1\in\mathbb{R}^{p_1}$ follows a slash distribution with location parameter $\bmu_1\in\mathbb{R}^{p_1}$, scale matrix $\bSigma_{11}\in\mathbb{R}^{p_1\times p_1}$ and $\nu>0$ degrees of freedom. On the other hand, the conditional pdf of $\X_2|(\X_1=\x_1)$ is given by
\begin{eqnarray*}
f_{\X_2|\X_1}(\x_2|\x_1) &=& \frac{f_{\X}(\x_1,\x_2)}{f_{\X_1}(\x_1)} = \frac{\nu}{f_{\X_1}(\x_1)}\int_0^1 u^{\nu-1} \phi_p\left(\x; \bmu,u^{-1}\bSigma\right) \dr u \\
&=& \frac{\nu}{f_{\X_1}(\x_1)}\int_0^1 u^{\nu-1} \phi_{p_1}\left(\x_1; \bmu_1,u^{-1}\bSigma_{11}\right) \phi_{p_2}\left(\x_2; \bmu_{2.1},u^{-1}\bSigma_{2.1}\right) \dr u. \hspace*{2cm}
\end{eqnarray*}
Then, it is possible to notice that the Slash distribution is not closed under conditioning. Furthermore, the pdf of $\X_2|(\X_1=\x_1)$ belongs to the elliptical family of distributions with dgf $g(t) = \int_0^1 u^{\nu+p/2-1} \exp\{-u(t + \delta_1(\x_1))/2\} \dr u$, i.e., $\X_2|(\X_1=\x_1) \sim \mbox{E}\ell(\bmu_{2.1}, \bSigma_{2.1}, \nu; g)$. To determine the mean of the random vector $\X_2|(\X_1=\x_1)$, we compute the conditional expected value of the $i$th element of $\X_2$ as follows
\begin{eqnarray*}
\E(\X_{2i}|\X_1=\x_1) &=& \int_{\mathbb{R}^{p_2}} x_{2i} f_{\X_2|\X_1}(\x_2|\x_1) \dr\x_2 \\
&=& \frac{\nu}{f_{\X_1}(\x_1)} \int_{\mathbb{R}^{p_2}} x_{2i} \int_0^1 u^{\nu-1} \phi_{p_1}\left(\x_1; \bmu_1,u^{-1}\bSigma_{11}\right) \phi_{p_2}\left(\x_2; \bmu_{2.1},u^{-1}\bSigma_{2.1}\right) \dr u\, \dr\x_2 \\
&\stackrel{\mbox{Fubini}}{=}& \frac{\nu}{f_{\X_1}(\x_1)} \int_0^1 u^{\nu-1} \phi_{p_1}\left(\x_1; \bmu_1,u^{-1}\bSigma_{11}\right) \int_{\mathbb{R}^{p_2}} x_{2i} \phi_{p_2}\left(\x_2; \bmu_{2.1},u^{-1}\bSigma_{2.1}\right) d\x_2 \, \dr u \\
&=& \frac{\mu_{2.1}^{(i)} \nu}{f_{\X_1}(\x_1)} \int_0^1 u^{\nu-1} \phi_{p_1}\left(\x_1; \bmu_1,u^{-1}\bSigma_{11}\right) \dr u = \mu_{2.1}^{(i)}, \quad \forall i, \nu>0
\end{eqnarray*}
where $\mu_{2.1}^{(i)}$ represents the $i$th element of the vector $\bmu_{2.1}$, and $\E(\X_2|\X_1=\x_1) = \bmu_{2.1}$. Now, to compute the elements of the variance-covariance matrix of the conditional random vector, we first determine $\E(X_{2i}X_{2j}|\X_1=\x_1)$ for all $i,j=1,\ldots,p_2$, as
\begin{eqnarray*}
\E(X_{2i}X_{2j}|\X_1=\x_1) &=& \int_{\mathbb{R}^{p_2}} x_{2i}x_{2j} f_{\X_2|\X_1}(\x_2|\x_1) d\x_2 \\
&=& \frac{\nu}{f_{\X_1}(\x_1)} \int_{\mathbb{R}^{p_2}} x_{2i} x_{2j} \int_0^1 u^{\nu-1} \phi_{p_1}\left(\x_1; \bmu_1,u^{-1}\bSigma_{11}\right) \phi_{p_2}\left(\x_2; \bmu_{2.1},u^{-1}\bSigma_{2.1}\right) \dr u \,\dr\x_2 \\
&\stackrel{\mbox{Fubini}}{=}& \frac{\nu}{f_{\X_1}(\x_1)} \int_0^1 u^{\nu-1} \phi_{p_1}\left(\x_1; \bmu_1,u^{-1}\bSigma_{11}\right) \int_{\mathbb{R}^{p_2}} x_{2i} x_{2j} \phi_{p_2}\left(\x_2; \bmu_{2.1},u^{-1}\bSigma_{2.1}\right) \dr\x_2 \,\dr u \\
&=& \frac{\nu}{f_{\X_1}(\x_1)} \int_0^1 u^{\nu-1} \phi_{p_1}\left(\x_1; \bmu_1,u^{-1}\bSigma_{11}\right) \left( u^{-1}\sigma_{2.1}^{(ij)} + \mu_{2.1}^{(i)}\mu_{2.1}^{(j)} \right) \dr u \\
&=& \frac{\sigma_{2.1}^{(ij)} \nu}{f_{\X_1}(\x_1)} \int_0^1 u^{\nu-2} \phi_{p_1}\left(\x_1; \bmu_1,u^{-1}\bSigma_{11}\right) \dr u + \mu_{2.1}^{(i)}\mu_{2.1}^{(j)} \\
&=& \frac{\nu}{\nu-1}\left(\frac{\mbox{SL}_{p_1}(\x_1; \bmu_1,\bSigma_{11}, \nu-1)}{\mbox{SL}_{p_1}(\x_1; \bmu_1,\bSigma_{11}, \nu)}\right) \sigma_{2.1}^{(ij)} + \mu_{2.1}^{(i)}\mu_{2.1}^{(j)}, \quad \nu>1,
\end{eqnarray*}
where $\sigma_{2.1}^{(ij)}$ is the $(i,j)$th element of the matrix $\bSigma_{2.1}$. From these results, we have that
\begin{eqnarray*}
\mathrm{Cov}(X_{2i}, X_{2j}|\X_1=\x_1) = \frac{\nu}{\nu-1}\left(\frac{\mbox{SL}_{p_1}(\x_1; \bmu_1,\bSigma_{11}, \nu-1)}{\mbox{SL}_{p_1}(\x_1; \bmu_1,\bSigma_{11}, \nu)}\right) \sigma_{2.1}^{(ij)}, \quad\nu>1.
\end{eqnarray*}
Therefore, the covariance matrix of the random vector $\X_2|(\X_1=\x_1)$ will be given by
\begin{eqnarray*}
\mathrm{Cov}(\X_2|\X_1=\x_1) = \frac{\nu}{\nu-1}\left(\frac{\mbox{SL}_{p_1}(\x_1; \bmu_1,\bSigma_{11}, \nu-1)}{\mbox{SL}_{p_1}(\x_1; \bmu_1,\bSigma_{11}, \nu)}\right) \bSigma_{2.1}, \quad \nu>1.
\end{eqnarray*}

\end{document}